\documentclass{aa}

\usepackage{graphicx}

\usepackage{txfonts}

\usepackage{hyperref}
\hypersetup{
     colorlinks = true,
     linkcolor = blue,
     anchorcolor = blue,
     citecolor = blue,
     filecolor = blue,
     urlcolor = blue
}

\usepackage{bm}
\usepackage{caption}
\usepackage{subcaption}
\usepackage{bigstrut}
\usepackage{xspace}

\mathchardef\mhyphen="2D

\defcitealias{Lissauer&Gavino_2021}{Paper I}
\defcitealias{Bartram+etal_2021}{Paper II}
\defcitealias{Obertas+etal_2017}{OVT17}
\defcitealias{Petit+etal_2020}{PPDJ20}

\begin{document}

   \title{Orbital stability of compact three-planet systems}

   \subtitle{III. The role of three-body resonances}

 \author {
       S. Gavino \inst{1}
        \and J. J. Lissauer \inst{2}
         }    

 \institute{
        Dipartimento di Fisica e Astronomia, Università di Bologna, Via Gobetti 93/2, 40122, Bologna, Italy; Niels Bohr Institute and Centre for Star and Planet Formation, University of Copenhagen, Øster Voldgade 5-7, 1350, Copenhagen K, Denmark
        \and Space Science \& Astrobiology Division, MS 245-3, NASA Ames Research Center, Moffett Field, CA 94035, USA
    }

\offprints{ Sacha Gavino\\
\email{sacha.gavino@unibo.it}}

   \date{Received October 2025; accepted 2026}

\abstract
{Observational surveys show that at least $\sim 30$\% of short-period multi-planet systems contain planets in tightly packed orbits. Some of these systems are locked in stable chains of mean-motion resonances. Although significant advances have been made in recent years, our understanding of the dynamical stability of these systems remains incomplete. Overall, numerical simulations show that the general trend is an exponential increase in a system's lifetime with increasing orbital separation, with mean motion resonances being a key mechanism that modulates stability. Tightly packed three-planet systems exhibit a specific dynamical behavior that is not yet found in systems with more planets: a small yet significant fraction of the phase space of very compact three-planet systems is anomalously stable.} 
{This study investigates the dynamics of extremely compact three-planet systems, focusing on those that are anomalously long-lived, and explores their connection to resonant chains observed in exoplanetary systems.} 
{We used numerical integrators to explore the dynamics of coplanar, initially circular, tightly packed, and equal-mass three-planet systems over timescales on the order of the star's lifetime, with a very high resolution in orbital separation. We analyzed the results in light of recent analytical studies.}
{We identify the fraction of phase space with anomalously stable, very tightly packed orbits, including systems that are orders of magnitude longer-lived than expected from the exponential trend. We show an unambiguous link between the stability of these systems and isolated three-body mean-motion resonances.}
{We find that very tightly packed three-planet systems can remain stable when captured in a small subset of isolated zeroth-order three-body resonances. Stability critically depends on the initial orbital longitudes and on the interplay between the three-body mean-motion resonance network and the two-body mean-motion resonance network.    
}


   \maketitle
%

\section{Introduction}\label{sec:intro}

Approximately 1000 multi-planet systems have been identified in recent decades, most by NASA’s {\it Kepler} mission \citep{Lissauer+etal_2011a, Lissauer+etal_2011b, Fabrycky+etal_2014, Lissauer+etal_2024} and Transiting Exoplanet Survey Satellite (TESS). Hundreds of these systems contain three or more super-Earth and/or sub-Neptune planets on nearly coplanar orbits in compact systems, and many individual systems show approximate uniformity in mass and orbital spacing \citep{Millholland+etal_2017, Weiss+etal_2018, Weiss+etal_2023}. The formation routes to such system architectures have recently been discussed \citep[e.g.,][]{Goldberg+Batygin_2022, Tejada_Arevalo+etal_2022}, but many questions remain unanswered. In particular, although simulations suggest that planets are typically captured in mean-motion resonance (MMR) chains due to planet-disk interactions, studies show that planetary systems can become unstable once the dissipative action of the disk stops, and observations indicate that the majority of tightly packed systems occupy orbits away from resonant configurations \citep[e.g.,][] {Fabrycky+etal_2014, Batygin+Adams_2017,Izidoro+etal_2017, Pichierri+Morbidelli_2020, Izidoro+etal_2021}. Indeed, observations show that the probability of a multi-planet system being locked in a resonant chain decreases with system age \citep{Dai_2024}. Nevertheless, even though the fraction of observed resonant chains among old planetary systems is small, understanding the conditions that ensure their stability may help explain why these resonant configurations are uncommon.

Analytical studies of two-planet systems have provided a satisfactory characterization of their dynamical stability \citep{Hill_1878a}. The stability criteria are mainly based on orbital spacing \citep[e.g.,][]{Gladman_1993} or on the overlap of MMRs \citep[e.g.,][]{Chirikov_1979, Wisdom_1980, Deck+etal_2013, Hadden+Lithwick_2018}.

However, deriving analytical stability criteria becomes significantly more challenging for systems containing three or more planets. Many authors have used numerical integrations to determine the mechanisms that lead to instability in these systems. In most cases, these studies are restricted to the idealized case of systems of equal-mass planets on equally spaced (in terms of period ratios of neighboring planets) orbits. Although these intensively studied systems are  ``idealized'', they provide an effective tool for deriving general dynamical trends in planetary systems.  \citet{Chambers+etal_1996} performed the first numerical investigation. Successive studies focused on specific numbers of planets \citep[e.g.,][]{Obertas+etal_2017, Lissauer&Gavino_2021}, different masses \citep[e.g.,][]{Smith+Lissauer_2009, Morrison+Kratter_2016, Rice+Steffen_2023}, initial orbital eccentricities \citep{Gratia+Lissauer_2021}, or systems with co-orbital planets \citep{Smith+Lissauer_2010}. All of these works show the general trend that a system's lifetime increases exponentially with increasing initial orbital spacing.

Other authors have derived analytical or semi-analytical solutions to predict the survival time of systems with more than two planets. In particular, \citet{Quillen_2011} introduced a resonance overlap criterion to estimate the role of three-body MMRs (3BRs) in causing destabilization in tightly packed multi-planet systems. More recently, \citet[henceforth \citetalias{Petit+etal_2020}]{Petit+etal_2020} presented an analytical model that estimates the lifetimes of compact systems based on the overlap of zeroth-order 3BRs. Subsequently, \citet{Petit+etal_2021} developed an integrable model for first-order 3BRs by generalizing a similar approach to that used for two-planet systems.

Systems with three planets represent about a quarter of all multi-planet systems in the {\it Kepler} planet candidate catalog \citep{Lissauer+etal_2024} and constitute a bridge between the dynamically well-characterized one- and two-planet systems and general multi-planet systems. Numerical simulations of three-planet systems (\citealt{Lissauer&Gavino_2021}; henceforth \citetalias{Lissauer&Gavino_2021}) show specific behavior that was not reported in numerical simulations with more planets (\citealt{Smith+Lissauer_2009}; \citealt{Obertas+etal_2017} hereafter \citetalias{Obertas+etal_2017}; \citealt{Outland+etal_2026}). 

This work extends \citetalias{Lissauer&Gavino_2021} and (\citealt{Bartram+etal_2021}; henceforth \citetalias{Bartram+etal_2021}) who quantitatively characterized the lifetime dependence on the orbital spacing of circular, tightly packed three-planet systems and explored the post-instability impact behavior as a function of the physical (not orbital) radii of the planets. As for systems with four or more planets, they show that three-planet systems follow a trend in which a system's lifetime increases approximately exponentially with the initial orbital separation, although three-planet systems tend to be
globally longer-lived, with the magnitude of the difference increasing at larger separations. More surprisingly, they show that a notable fraction of very tightly packed three-planet systems exhibit peculiar behavior due to specific pairings of orbital separation and initial longitudes, resulting in anomalous lifetimes that are multiple orders of magnitude longer than the exponential trend. Although such systems likely constitute only a tiny fraction of all planetary systems in the galaxy, detailed investigation is warranted, as their stability mechanisms may share features with observed tightly packed resonant chains that remained stable after the dissipation of the protoplanetary disk gas. This work extends the numerical investigation of anomalously stable three-planet systems initiated by \citetalias{Lissauer&Gavino_2021}, revealing links to 3BRs.

\section{Numerical methods}\label{sec:method}

\subsection{Integration packages and duration}

We performed all integrations presented in Sects.\,\ref{sec:random} and \ref{sec:restricted} using the Wisdom-Holman mixed-variable symplectic (MVS) method embedded in the Mercury software package \citep{Chambers_1999}, with a time step of 18 days. We considered a system unstable when two orbits cross (more precisely, when the apoapsis of the innermost orbit becomes larger than the periapsis of the middle orbit, or the apoapsis of the middle orbit becomes larger than the periapsis of the outermost orbit). We integrated each system  until it became unstable or 10$^{10}$ initial orbital periods of the innermost planet elapsed. We defined the time that elapses until the system becomes unstable or reaches the maximum integration time as the system lifetime; for systems that time out, this provides only a lower bound of 10$^{10}$ periods. We denote the system lifetime as $t_\mathrm{c}/t_\mathrm{0}$, where $t_\mathrm{0}$ is the period of the innermost planet (1 yr). 

In Sects.\,\ref{sec:spikes}, and \ref{sec:nonequally}, we used WHFast, a fast implementation of a Wisdom-Holman symplectic integrator \citep{reboundwhfast,wh} in the N-body code REBOUND \citep{rebound} to perform further analyses of a selection of systems of interest with the same timestep of 18 days.

\subsection{Initial orbital configurations}

In this study, all systems have three planets of equal mass, with each planet's mass equal to that of the Earth ($M_{j} = 1$~M$_\oplus$ for $j=1,2,3$), orbiting a one-solar-mass star $M_\star=1$~M$_\odot$. The systems are coplanar, and the orbits of the planets are in the same direction and initially circular. The innermost planet orbits at 1 AU from the star. 

Although the Hill radius is not the most appropriate measure of orbital separation (stability is better described by a spacing scaled as the planet-to-star mass ratio to the power of 1/4; \citetalias{Petit+etal_2020}), we used it for consistency with previous studies. The mutual Hill radius of neighboring planets is defined as

\begin{equation}
\label{eq:Hillradius}
        R_{H_{j, j+1}} \equiv X \left(a_j + a_{j+1}\right) ,
\end{equation} 

\noindent where $a_\mathrm{j}$ and $a_\mathrm{j+1}$ are the semi-major axes of the $j$th and $(j+1)$th planets, respectively, and where $X$ is

\begin{equation}
\label{eq:X}
       X \equiv \frac{1}{2} \left[\frac{2{\rm M}_\oplus}{{3({\rm M}_\odot}+(j-1){\rm M}_\oplus)} \right]^\frac{1}{3}.
\end{equation}

The relationship between the semi-major axes of successive pairs of planets can be expressed in terms of mutual Hill radius as follows:

\begin{equation}
\label{eq:beta}
       a_{j+1} = a_j + \beta R_{H_{i,j+1}},
\end{equation}

\noindent where we refer to  $\beta$ as the dynamical separation. Equal values of $\beta$ do not imply equal orbital distances, even for equal-mass planets. A system of two planets that are initially on circular and coplanar orbits is stable if $\beta$ exceeds $\approx 2 \sqrt{3}$ \citep{Gladman_1993}. This system is Hill stable. In the case of Earth-mass planets, this corresponds to an orbital period ratio of $\sim$ 1.0680.

Similar to \citet{Obertas+etal_2017}, we used Eq.\,\ref{eq:X} to express the semi-major axes of successive pairs of planets of equal mass as

\begin{equation}
\label{eq:beta2}
       a_{j+1} = a_j \bigg( \frac{1 + \beta X}{1 - \beta X}\bigg).
\end{equation}

\noindent Using Kepler's third law, this allows us to express the period ratios of successive pairs as a function of orbital separation,

\begin{equation}
\label{eq:periodratio_beta}
       \frac{P_{j+1}}{P_j} = \bigg( \frac{1 + \beta X}{1 - \beta X}\bigg)^\frac{3}{2}.
\end{equation}

We considered four sets of initial angular orbital elements, with coordinates such that the innermost planet begins at 0 longitude; the sets differ only by the longitudes of the middle and outer planets. We summarize these sets of longitudes in Table\,\ref{tab:sets}. In the first set (random), we randomly and independently drew the initial longitudes of the middle and outer planets, $\lambda_{2}$ and $\lambda_{3}$, from a uniform distribution. This method is similar to those used, for example, in \citet{Obertas+etal_2017} and in Sect. 3 of \citetalias{Lissauer&Gavino_2021}. We present lifetimes of these systems in Sect. \ref{sec:random:tc}. 
 
The second, third, and fourth sets--restricted random (RR), specific 1 (S1), and specific 2 (S2), respectively--have restricted initial longitudes for the middle and outer planets. We give the rationale for choosing restricted longitudes in the following sections.

\setlength{\tabcolsep}{4pt}
\begin{table}
\centering
\caption{Initial planetary longitudes.$^a$ \label{tab:sets}}

\begin{tabular}{p{2.5cm} p{1.6cm} c c}
\hline
Sets & Section & $\lambda_2$ & $\lambda_3$ \\
\hline
Random & \ref{sec:random} & random & random \\
Restricted Random (RR) & \ref{sec:restricted} & random & $-2^\circ \leq \lambda_2 \leq 2^\circ$ \\
Specific \#1 (S1) & \ref{sec:restricted} & 166.273784$^\circ$ & 165.110014$^\circ$ \\
Specific \#2 (S2) & \ref{sec:restricted} & 207.16381$^\circ$ & 208.66160$^\circ$ \\
\hline
\end{tabular}

\medskip
\tablefoot{$^a$ Longitude values are given to input precision.}

\end{table}

\section{Results: Systems with random initial longitudes}  \label{sec:random}

\subsection{Lifetime versus orbital separation} \label{sec:random:tc}

In this section, we extend the work of \citetalias{Lissauer&Gavino_2021} to better characterize anomalously long-lived systems and their dependence on the initial longitudes. To detect previously unidentified anomalous systems, we performed a blind search by integrating systems in the random set with a resolution of $5\times10^{6}$ per unit $\beta$ (more than three orders of magnitude finer than in \citetalias{Lissauer&Gavino_2021}) over the range [2$\sqrt{3}$, 5.00] and a resolution of $2\times10^{4}$ within the range (5.00, 5.60). The entire random set comprises more than 7,500,000 integrated systems. 

\begin{figure*}
    \centering
    \begin{subfigure}{0.49\linewidth}
        \includegraphics[width=\linewidth]{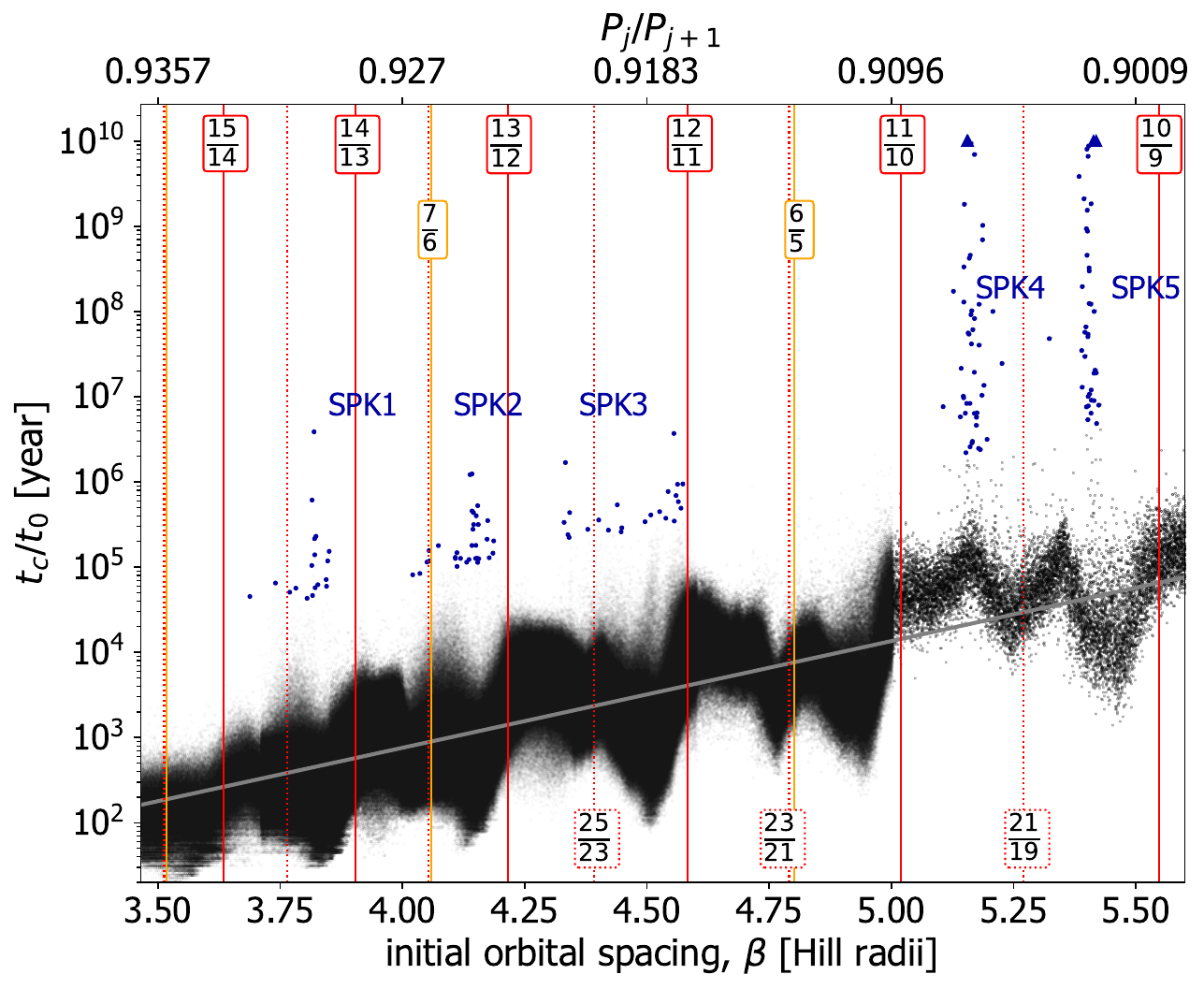}
        \caption{}
        \label{fig:drawingA}
    \end{subfigure}
    \begin{subfigure}{0.49\linewidth}
        \includegraphics[width=\linewidth]{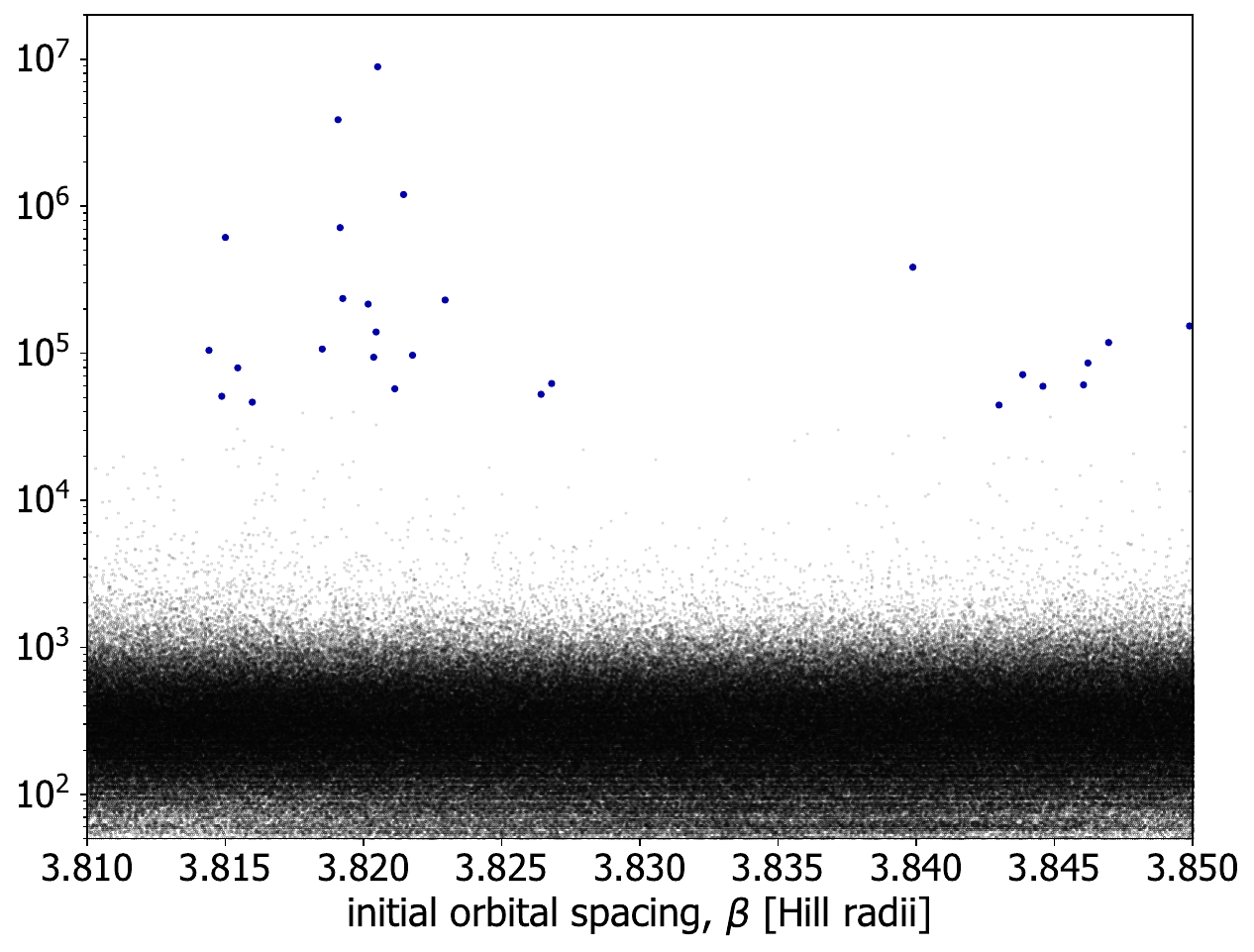}
        \caption{}
        \label{fig:drawingB}
    \end{subfigure}
    \caption{Lifetime, $t_c$, of three-planet systems as a function of initial separation
of neighboring planets' orbits (in units of $\beta$) for the Random set, where the initial longitudes of the middle and outer
planets are selected randomly. Panel (a) shows the full view with a density of $5\times10^{6}$ per unit $\beta$ in the range [2$\sqrt{3}$, 5.00], and a density of $2\times10^{4}$ in the range (5.00, 5.60]. Systems with a lifetime more than 5$\sigma$ above the exponential fit are shown with larger navy-blue dots. Upward-pointing triangles represent systems that remain stable for the entire elapsed time, 10$^{10}$ years. The upper horizontal axis shows the initial period ratios of neighboring planets, as defined by Eq.\,\ref{eq:periodratio_beta}. Solid (dotted) vertical red lines denote the locations of first-order (second-order) 2BRs between adjacent planet pairs, and the solid vertical gold lines denote first-order 2BRs
between the innermost and outermost planets. Solid gray line segments represents the exponential fit to all points with a density of 20,000 per unit $\beta$ in the range [3.4645, 5.6000]. The terms SPK\textit{i} denote the \textit{i}th spike with $i \in$ [1,2,3,4,5]. Panel (b) shows a zoomed-in view with a density of $10^{7}$ in the range $\beta \in$~[3.81, 3.85].}
    \label{fig:drawing}
\end{figure*}

Figure \ref{fig:drawingA}  shows the lifetime as a function of the initial separation for all integrated systems in the random set. As other studies show, the systems' lifetimes, $t_c$, approximately follow a linear trend with the initial separation measured in $\beta' \equiv \beta - 2\sqrt3$, given by $\log~t_c = b' \beta' + c'$, where $b'$ and $c'$ are constants \citep[e.g.,][]{Chambers+etal_1996, Smith+Lissauer_2009, Obertas+etal_2017,Quarles_2018} and \citetalias{Lissauer&Gavino_2021}. We fit our data covering the range $(2\sqrt{3}, 5.6]$ in $\beta$ using a uniform subset with a resolution of $2\times10^4$, which yields $b' = 1.25212$ and $c' = 2.20944$. We define a system as anomalously long-lived when its lifetime is greater than 5$\sigma$ above the exponential fit, which corresponds to roughly two orders of magnitude larger than the exponential fit lifetime at the system's orbital separation.

\subsection{Anomalous systems} \label{sec:random:anom}

Fig.\,\ref{fig:drawing} clearly shows spikes corresponding to anomalously long-lived systems, with five successive spikes identifiable. As shown in later sections, these long-lived systems are associated with capture into specific 3BRs, which constitute the primary dynamical mechanism responsible for their stability.
We designate the first three spikes, from left to right in Fig.\,\ref{fig:drawingA},  as SPK1, SPK2, and SPK3. \citetalias{Lissauer&Gavino_2021} did not detect these spikes due to the lower resolution in $\beta$. We further increased the resolution by a factor of two in the SPK1 region (within the range [3.81, 3.85]) to better visualize the inner structures (Fig.\,\ref{fig:drawingB}). 

Two outer spikes, SPK4 and SPK5, appear surrounding $\beta \sim 5.15$ and $\beta \sim 5.40$, respectively, which appear to be longer-lived relative to the exponential fit than the three inner spikes. More significantly, they occupy a much larger fraction of phase space, since they have more points despite a resolution of 250 or 500 times coarser. Four systems survived for ten billion years in SPK4 and SPK5, and have values $\beta$ = 5.1547, 5.15665, 5.4135, and 5.41885. This means that using specific initial longitudes can allow the systems to survive for more than six orders of magnitude longer than the lifetime approximated by the exponential fit. Note that these two spikes were already found by \citetalias{Lissauer&Gavino_2021}.

The spikes are not uniformly spaced along the horizontal axis in Fig.~\ref{fig:drawing}. In particular, we can see a wide region between SPK3 and SPK4 where no spikes were found, which corresponds to a gap of approximately 0.60 in $\beta$ (between $\beta \approx 4.55$ and $\beta \approx 5.15$). This suggests that the inner and outer spikes may originate from different mechanisms. The origin of this wide region will be discussed in Sect. \ref{sec:gap}.

\begin{figure} 
\centering
\includegraphics[width=1.0\linewidth]{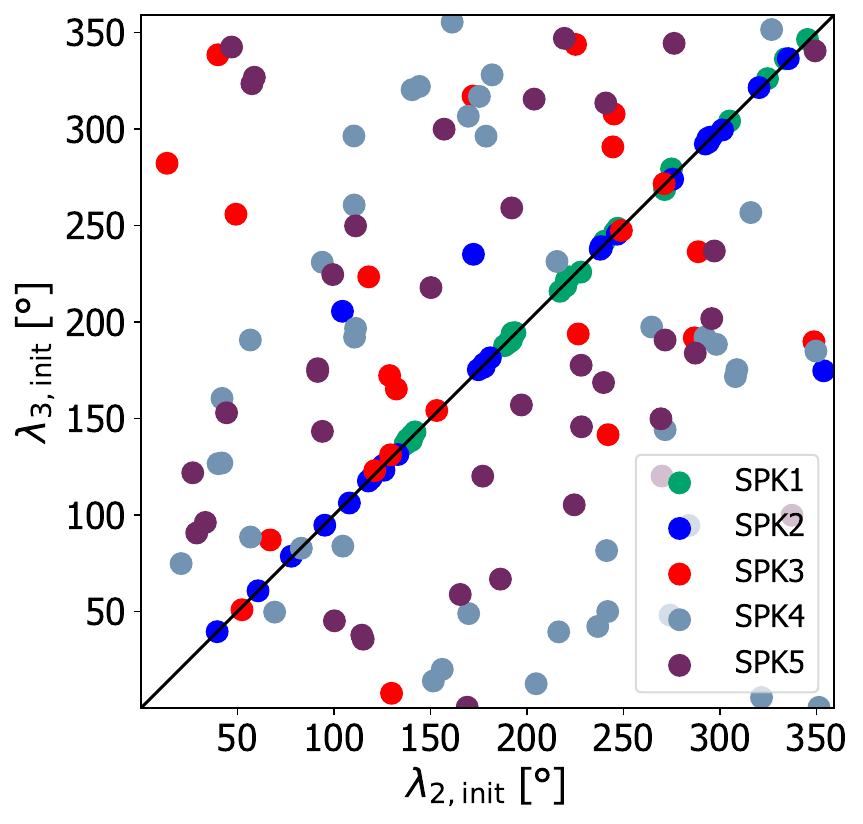}
\caption{Initial longitude of the middle planet versus the outer planet for systems from the random set whose lifetimes are more than 5$\sigma$ above the exponential fit (anomalous systems), i.e., systems in SPK1, SPK2, SPK3, SPK4, and SPK5. For SPK1, we include systems from the higher resolution batch shown in Fig.\,\ref{fig:drawingB}. The solid diagonal black curve marks the locus of points where the outer pair of planets start in conjunction. \label{fig:anomalous} }
\end{figure}

Similarly to \citetalias{Lissauer&Gavino_2021}, we find that specific combinations of initial conditions lead to anomalous long-term stability. In this work, we detect three additional inner spikes: SPK1, SPK2, and SPK3. Fig. \ref{fig:anomalous} shows the distribution of the spike systems as a function of the initial angles.

For the three inner spikes in Fig.\,\ref{fig:drawing}, we find 18 (34 including the higher resolution shown in Fig.\,\ref{fig:drawingB}), 33, and 23 systems in SPK1, SPK2, SPK3, respectively. The figure shows a clear correlation between the initial conjunction of the outer pair of planets and dynamical stability. In total, 34, 30, and seven systems, in SPK1, SPK2, and SPK3, respectively, start near the curve $\lambda_2 = \lambda_3$ ($\pm 2$°), which corresponds to 100$\%$, 91$\%$, and 30$\%$ of all systems in each spike, respectively. This result should be interpreted with the caveat discussed in Sect. \ref{sec:spikes123}.

We find 43 systems in SPK4 and 39 in SPK5, despite the much coarser resolution of our simulations for these less closely spaced systems. We do not observe clear angle preferences for SPK4 and SPK5, although the small number of systems may prevent statistical trends from being immediately visible. \citetalias{Lissauer&Gavino_2021} show a correlation between initial angles and the stability of the systems in SPK4 and SPK5. We discuss this correlation in Sect. \ref{sec:spikes45}. Fig.\,\ref{fig:anomalous}  shows no evident correlation between the lifetime of the system and other conjunctions (of the inner pair or between the first and third planets). 

\section{Lifetimes of the restricted and specific sets} \label{sec:restricted}

\begin{figure} 
\centering
\includegraphics[width=1.0\linewidth]{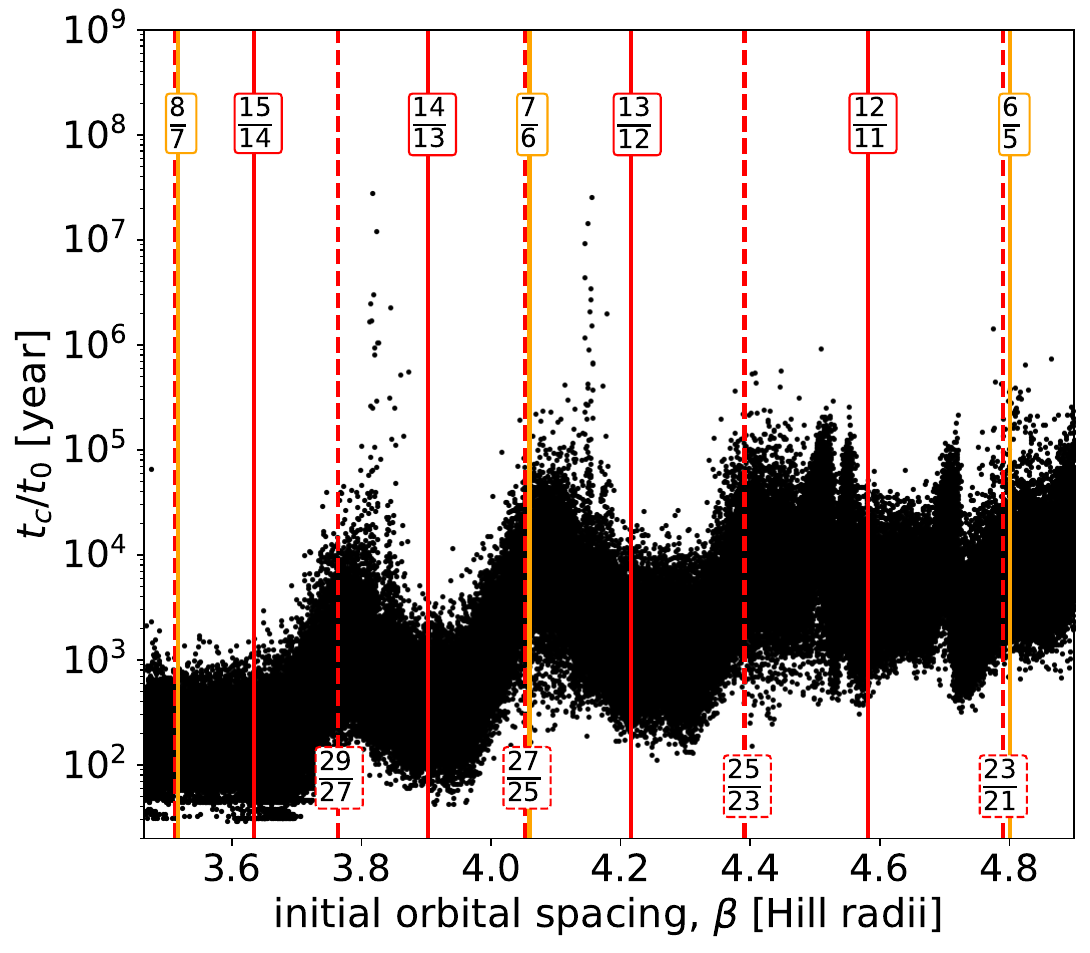}
\caption{Lifetime, $t_c$, of three-planet systems as a function of the initial separation
of neighboring planets' orbits for the RR set (see Table\,\ref{tab:sets}), with a fixed density of $2\times10^5$ per unit of $\beta$ over the range [2$\sqrt{3}$, 4.9000]. \label{fig:restr} }
\end{figure}

To determine how close-to-conjunction initial configurations affect the ``strength'' and shape of the spikes, we simulated an additional set of systems, called the RR set, in which the initial longitude of the middle planet, $\lambda_{2}$, was randomly selected and the initial longitude of the outer planet, $\lambda_{3}$, was randomly drawn from a range within 2° around $\lambda_{2}$. We restricted the simulations inside $\beta = 4.9$, since only SPK1, SPK2, and SPK3 show a correlation with the initial conjunction, and we used a resolution of $10^{-5}$ in $\beta$. Fig. \ref{fig:restr} shows the results. Our simulations reproduce the equivalents of SPK1 and SPK2, but they are thinner and longer-lived. By contrast, SPK3 does not appear to change significantly. Combining SPK1 and SPK2, 20 systems survive for more than 10$^6$ years and four systems survive for more than 10$^7$ (two in each spike). There is also a notable outlier very close to the critical $\beta$ value for two-planet systems (2$\sqrt{3}$), which survives for about $8\times10^4$ years. 

\begin{figure*} 
\centering
\includegraphics[width=1.0\linewidth]{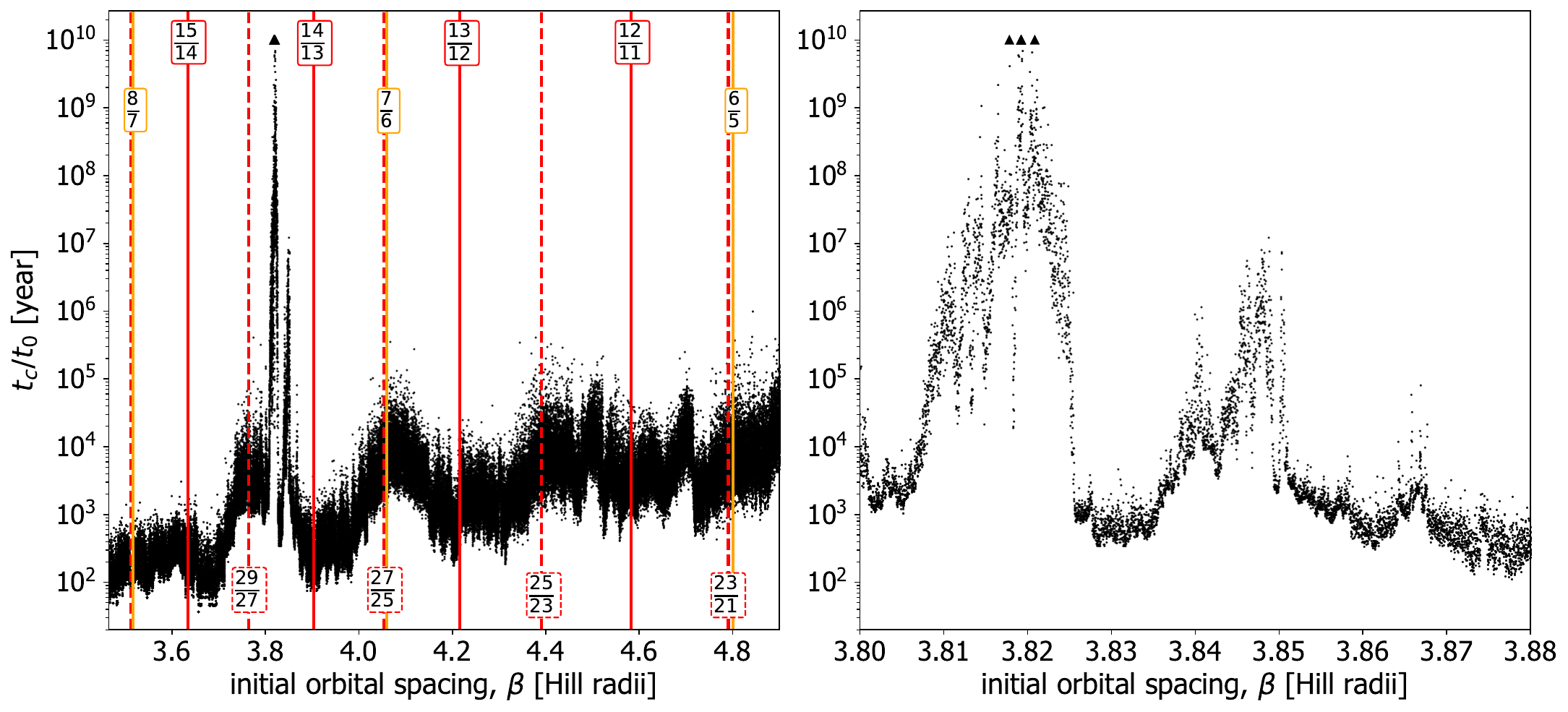}
\caption{Left: lifetime, $t_c/t_0$, of three-planet systems as a function of the initial separation
of neighboring planets' orbits for the set S1 ($\lambda_\mathrm{2,init}$ = 166.273784° and $\lambda_\mathrm{3,init}$ = 165.110014°). Triangles
represent systems that remain stable for the entire time interval, 10$^{10}$ years. Five systems survive for ten billion years; they have
initial orbital separations of $\beta$ = 3.81783, 3.81921, 3.81922, 3.81926, and 3.82086.  The resolution is 10$^5$ per unit of $\beta$. Right: Zoomed-in view of the spike in the range [3.80, 3.87].  \label{fig:special1} }
\end{figure*}

From the RR set, the longest-lived system in SPK1 has $\beta = 3.81563$, a lifetime of $2.72\times10^7$ years, and initial angles $\lambda_2 = 166.273784$° and $\lambda_3 = 165.110014$°. We further restricted the angle dependence on the lifetime in SPK1, by performing an additional set of simulations, the S1 set, using the system's initial longitude values. All systems in this set start with these initial angles and use the same resolution and $\beta$ range as the RR set. The left panel of Fig. \ref{fig:special1} shows the results. In this set, SPK1 is more prominent and longer-lived than in the other sets. A total of five systems remain stable for $10^{10}$ years. The spike also exhibits a complex structure (Fig.~\ref{fig:special1} right), composed of many successive sub-dips and sub-spikes. In particular, the spike comprises two major sub-spikes, one longer-lived than the other.

\begin{figure} 
\centering
\includegraphics[width=1.0\linewidth]{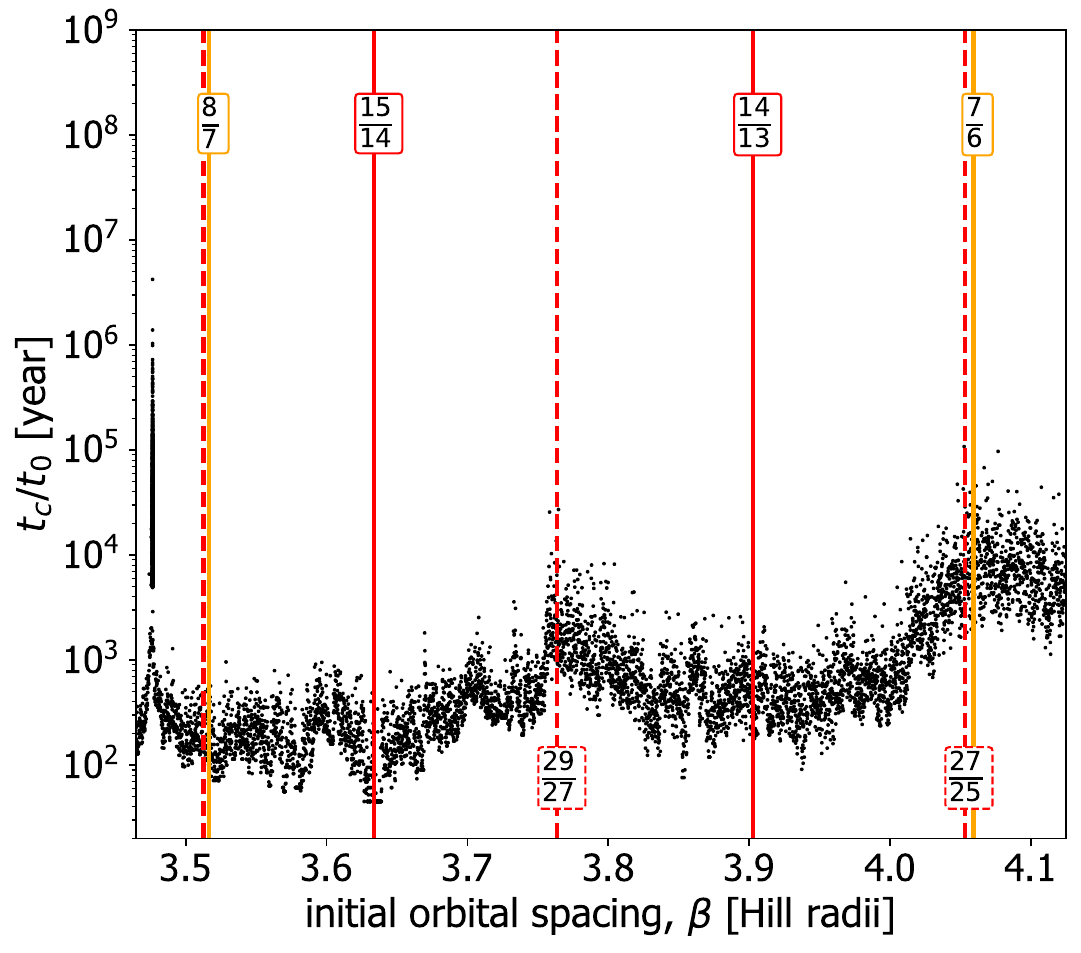}
\caption{Lifetime, $t_c$, of three-planet systems as a function of the initial separation
of neighboring planets' orbits for the set S2 ($\lambda_\mathrm{2,init}$ = 207.16381° and $\lambda_\mathrm{3,init}$ = 208.66160°). The resolution is 10$^6$ per unit $\beta$ in the range [3.47643,  3.47645] and $10^4$ elsewhere in the range [3.4645, 4.1300]. \label{fig:special2} }
\end{figure}

As mentioned above, we observe an anomalously stable system very close to the critical $\beta$ value for two planets in the RR set (Fig.\,\ref{fig:restr}). We performed another set of simulations, the S2 set, to determine whether this system belongs to a spike and whether a spike can exist this close to the critical value. All systems in S2 start with the initial longitude values of this anomalous system: $\lambda_2 = 207.16381$° and $\lambda_3 = 208.66160$°. Figure\,\ref{fig:special2} confirms the presence of an extremely thin spike around $\beta=3.4744$, which is approximately 0.35\% wider than the critical $\beta$ value for two planets.

\section{Geometry of three-body resonances}\label{sec:3br}

The spikes identified in Sect.\,\ref{sec:random} do not appear to be related to any strong two-body mean motion resonances (2BRs). Another  possible mechanism for the anomalous stability of these long-lived systems is three-body mean motion resonances (in principle, secular resonances could also be responsible. \citepalias{Petit+etal_2020} predict that, within what they define as the overlap limit of the three-body resonances, the dynamics of a three-planet system (not close to any strong two-body mean motion resonances) is driven by Chirikov diffusion \citep{Chirikov_1979}, which quickly leads the system on a chaotic path perpendicular to the resonance network. If systems within these long-lived spikes are stabilized by relatively strong three-body resonances, the presence of such systems with very small orbital period ratios ( well inside the theoretical stability limit) requires further analysis.

This section provides the material needed to address the points discussed in the following sections. Previous studies provide more thorough and detailed descriptions \citep[e.g.,][]{Chirikov_1979, Aksnes_1988, Quillen_2011, Petit+etal_2020}.

In the following, we denote the mean motion of the $j$th planet in a system by $n_j$, where $j=1,2,3$ and  $n_1 > n_2 > n_3$. A system is in a zeroth-order 3BR if the orbital frequencies of the three bodies are linked by the relationship

\begin{equation}
\label{eq:P3br}
        \frac{p}{n_2 - n_3} = \frac{q}{n_1 -n_2} = \frac{p+q}{n_1 -n_3} \approx \frac{P_\mathrm{3br}}{2\pi}, 
\end{equation}
 
\noindent where $p$ and $q$ are positive integers. The recurrence period $P_\mathrm{3br}$ is the interval over which the three planets return to the same initial orbital configuration (this periodic configuration does not necessarily occur at the same longitude in the inertial frame). Multiplying the two leftmost equations  of Eq.\,\ref{eq:P3br} by the denominators of both fractions and rearranging the terms yields

\begin{equation}
\label{eq:3body}
       pn_1 - (p+q)n_2 + qn_3 = 0.
\end{equation}

\noindent We can integrate Eq.~\ref{eq:3body} to show that a relationship associated with the three-body resonance also exists between the three longitudes around a resonant angle $\phi$:  

\begin{equation}
\label{eq:3body_long}
       p\lambda_1 - (p+q)\lambda_2 + q\lambda_3 = \phi,
\end{equation}

\noindent where $\lambda_1$, $\lambda_2$, and $\lambda_3$ are the longitude of planets 1, 2, and 3, respectively. The system is in a 3BR if the resonant angle $\phi$ librates.

Following the approach of \citetalias{Petit+etal_2020}, we divided Eq.\,\ref{eq:3body} by $n_2$ and isolated the new term $n_3/n_2$. The ratio of mean motions relates to the ratio of the orbital periods as

\begin{equation}
\label{eq:ratios}
      P_j/P_{j+1} = n_{j+1}/n_j,
\end{equation}

\noindent so that Eq.\,\ref{eq:3body} becomes 

\begin{equation}
\label{eq:pratio}
      \frac{P_2}{P_3} = \frac{p}{q}\bigg( 1 - \bigg(\frac{P_1}{P_2}\bigg)^{-1}\bigg) + 1.
\end{equation}

The geometry of the zeroth-order 3BRs can therefore be described in a 2D plane defined by the period ratios of the inner and outer pairs (zeroth-order 3BRs are independent of the longitude of the periapse).
We express the ratio of the three-body resonant period, $P_\mathrm{3br}$, to the period of the inner planet, $P_1$,  in terms of the period ratios of the outer pair only, using Eq.\,\ref{eq:P3br} and Eq.\,\ref{eq:pratio}:

\begin{equation}
\label{eq:P3br_ratio}
       P_\mathrm{3br} = q P_1 \bigg[1 + \frac{q}{p}\big(1- \frac{P_2}{P_3}\big)\bigg]\bigg[\frac{q}{p}\big(1 - \frac{P_2}{P_3}\big)\bigg]^{-1}.
\end{equation}

\noindent The smallest possible value of $P_\mathrm{3br}$ is for $q/p=1$.

\begin{figure} 
\centering
\includegraphics[width=1.0\linewidth]{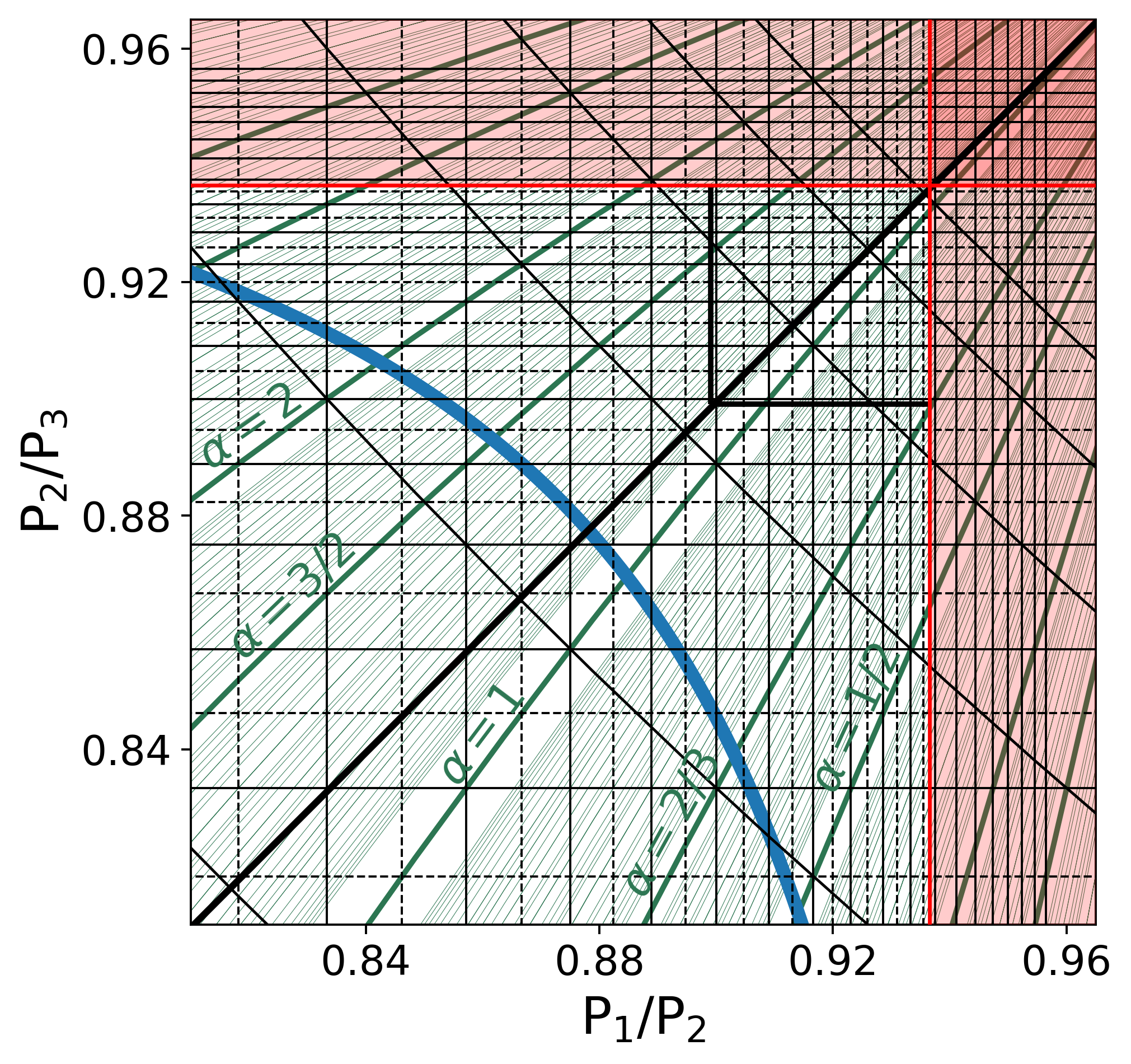}
\caption{Resonance locations and stability boundaries in the period-ratio plane. Red lines and shaded areas indicate the two-planet Hill-stability limits for circular orbits \citep{Gladman_1993, Petit+etal_2018}. The solid (dashed) black lines indicate the network of first-order (second-order) two-body MMRs. The bold solid  diagonal black line marks the locus of equal orbital period ratios (main diagonal). The oblique green lines represent the three-body resonance network for $p+q < 30$. Bold green lines indicate the loci of the three-body resonances for $\alpha = q/p = 1/2; 2/3; 1; 3/2; 2$. The blue curve indicates the theoretical overlap limit as defined by \citetalias{Petit+etal_2020} for Earth-mass planets. The black square contains the area investigated in this study, which spans inverse period-ratio values between $\sim 0.90$ and $\sim 0.94$. \label{fig:main_plane} }
\end{figure}

Throughout, we define $\alpha \equiv q/p$, where $p$ and $q$ are coprime. The indicated fraction $q/p$ always shows the lowest terms, but it also represents all other resonances with higher terms, since all resonances with the same $\alpha$ lie on top of each other. We denote the three-body resonances with $\alpha = w$ as 3BR$\alpha w$. Figure~\ref{fig:main_plane} shows the plane of the 2D period ratio. The resonances 3BR$\alpha1/2$, 3BR$\alpha2/3$, 3BR$\alpha1$, 3BR$\alpha3/2$, and 3BR$\alpha2$ are colored bold green because they are particularly important for the discussion. Using the density of the 3BR network, which is characterized by a filling factor, \citetalias{Petit+etal_2020} define a theoretical stability limit for a triplet of planets based on the overlap of 3BRs in the period ratio plane (their Eqs. 58 and 59). The limit is represented by the blue curve in Fig.\,\ref{fig:main_plane}. Inside this curve, the overlap of zeroth-order 3BRs is such that the network fully fills the space. For Earth-mass planets, the limit is at $P_{j+1}/P_j \approx 1.15$ for uniformly spaced orbits. Figure \,\ref{fig:main_plane} shows that the spike systems identified in Sect.\,\ref{sec:random} lie well within the theoretical stability limit, with the black square roughly indicating the boundaries of our numerical work space. We refer to the solid diagonal black line in Fig.\,\ref{fig:main_plane} as the ``main diagonal.'' This line represents the locus of systems in which two adjacent pairs have equal period ratios (i.e., geometrically spaced orbits). Therefore, all systems shown in Sects.\,\ref{sec:random} and \ref{sec:restricted}, as well as those in \citetalias{Lissauer&Gavino_2021} and \citetalias{Bartram+etal_2021}, start on the main diagonal.

Analogously, a pair of planets is in a two-body mean motion resonance (hereafter 2BR) if the longitudes are linked by the following relationship:

 \begin{equation}
\label{eq:2body}
       (p^\prime+q^\prime)\lambda_{j+1} - p^\prime \lambda_j + q^\prime \bar{\omega} = \phi_{2b}.
\end{equation}

\noindent where $p^\prime$ and $q^\prime$ are  positive integers and $\bar{\omega}$ is the pericenter longitude of the $j$th or $(j+1)$th planet. The integer $q^\prime =1$ for first-order 2BRs, $q^\prime =2$ for second-order 2BRs, and so on. A triplet of planets with a librating 3BR angle but without librating 2BR angles is in a pure 3BR.

An important aspect is the relationship between 3BRs and 2BRs. Consider a system locked in a 3BR with integers $p$ and $q$. If the inner pair is captured in a 2BR (of any order), we can express $P_2/P_1$ as a ratio $p'+q'/p'$ and Eq.\,\ref{eq:pratio} can be rewritten as  

\begin{equation}
\label{eq:correl_res3}
       \frac{P_2}{P_3} = 1 - \frac{p}{q} \frac{q'}{p'}.
\end{equation}

\noindent Equation \,\ref{eq:correl_res3} shows that if a system is captured in 3BR$\alpha1$, $P_3/P_2$ is necessarily a ratio of integers $(p''+q'')/p''$ such that $q'' = q'$. Thus, when 3BR$\alpha1$ intersects a $k$th-order 2BR of a pair of adjacent planets, the other pair is also systematically in a $k$th-order 2BR at the intersection. This behavior is specific to 3BR$\alpha1$, although other 3BRs also have a similar periodicity of 2BR crossings. For example, 3BR$\alpha2$ alternates between intersections of 2BRs of different order and 2BRs of the same order (see Fig.\,\ref{fig:main_plane}). More generally, all resonances with integers $p$ and $q$ that yield the same ratio $p/q$ lie on top of each other (i.e., when $p$ and $q$ are not coprime). The more 3BRs lie on top of each other, the higher the density of 2BR intersection crossings and the more isolated they are from the rest of the network. The 3BR$\alpha1$ exhibits the largest density of intersection crossings in the plane and is the most isolated of the 3BR network. The 3BR$\alpha2$, 3BR$\alpha3/2$, and 3BR$\alpha2/3$ resonances are also among the most isolated, although their degree of isolation decreases progressively as $\alpha$ departs from unity. These resonances therefore play an important role in stabilizing triplets of planets, particularly in regions where the 3BR network is dense, because systems near these resonances are more likely to remain safe from Chirikov diffusion and to be driven by 2BRs.  Appendix \ref{app:geo} provides more details on the geometry of three-body resonances.

\section{Dynamics of anomalously long-lived systems} \label{sec:spikes}

\subsection{Spikes SPK4 and SPK5} \label{sec:spikes45}

Figure \ref{fig:anomalous} shows no straightforward correlations between the lifetime of the spike systems and the initial angles for SKP4 and SKP5. \citetalias{Lissauer&Gavino_2021} did find a correlation between initial angular orbital elements and the survivability of spike systems with $\beta > 5$. We improved the statistics by exploring the dependence of initial angles on the survivability for a selection of systems in SPK4 and SPK5, rather than further increasing the resolution in $\beta$ in the random set around the spikes. We selected the first survivor ($t_c/t_0 \geq 10^{10}$ years) in SPK4, that is, the system with initial $\beta = 5.1547$ (hereafter called $b515$), as well as the system with initial $\beta = 5.4135$ (hereafter called $b541$). These systems have initial inverse period ratios of 0.9068 and 0.9023, respectively.

We first tested the lifetime dependence of SPK5 on the initial longitudes by exploring the lifetime of the system $b541$ for different sets of initial angles ($\lambda_2$ and $\lambda_3$). Because SPK5 is located in a dip (see Fig.~\ref{fig:drawing}), the non-long-lived systems will be short-lived (the dip reaches 2$\sigma$ below the exponential fit at $b541$, which corresponds to $t_c/t_0 \approx 4000$ yrs). We therefore consider a relatively high resolution in initial angles while keeping the total integration time relatively short. We adopted a resolution of 1° in initial longitude from 0 to 359°.

\begin{figure} 
\centering
\includegraphics[width=1.0\linewidth]{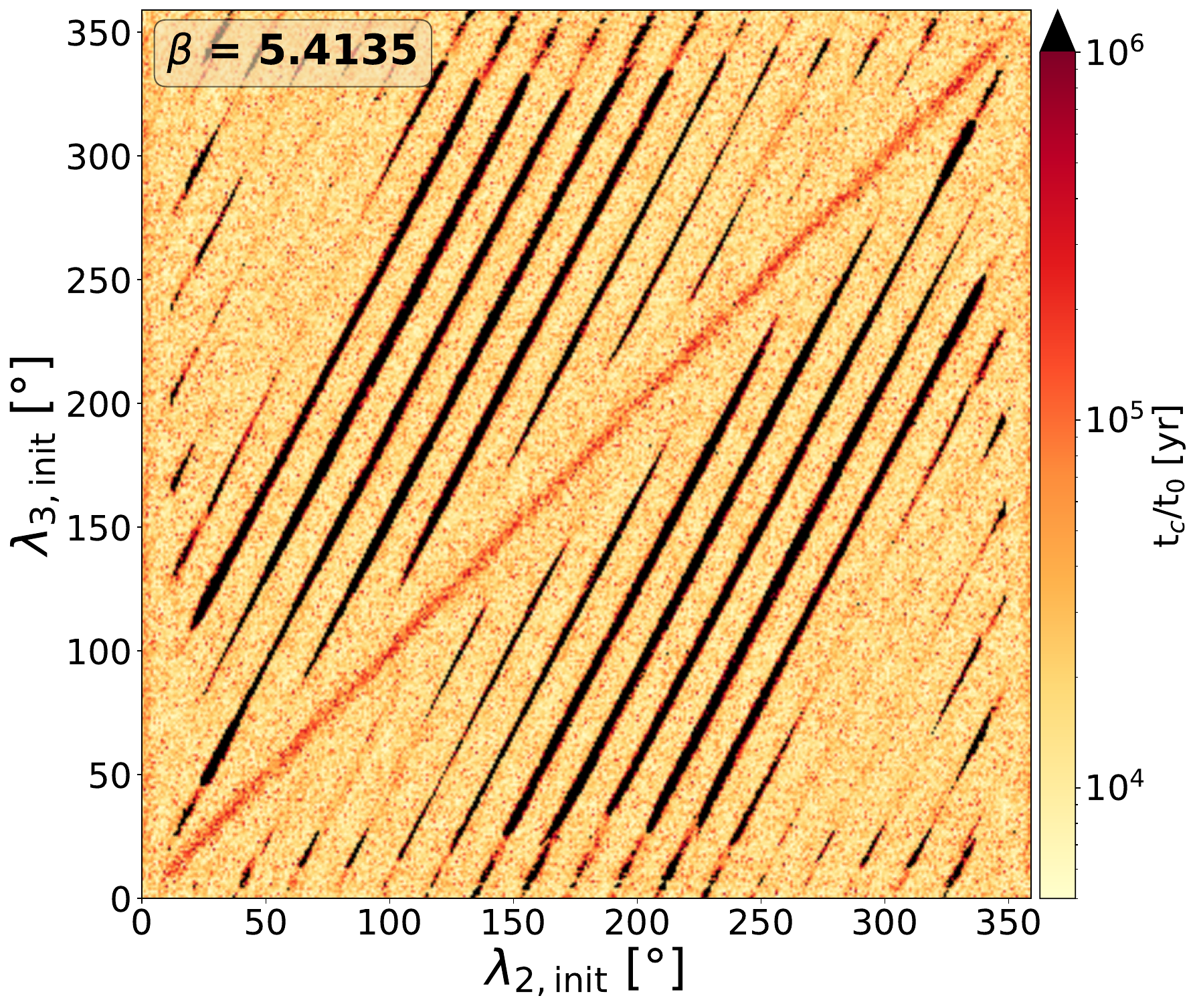}
\caption{System lifetimes as a function of initial longitudes of the middle planet ($\lambda_2$) and outer planet ($\lambda_3$). We selected the initial orbital separation $\beta = 5.4135$ from SPK5, where integration of our random set with initial longitudes $\lambda_2 = 228.228024$°   and  $\lambda_3 = 145.726985$° resulted in a lifetime > 10$^{10}$ years. The grid's resolution is 1° and it extends over the entire range, yielding $360\times360 = 129,600$ simulations. Integrations were halted after 10$^{6}$ yrs, with 12,001 runs reaching this time limit. These systems are shown in black.  \label{fig:stripes54135} }
\end{figure}

The resulting grid is shown in Fig. \ref{fig:stripes54135}, where a clear pattern of successive stripes of long-lived systems is visible. The stripes show gaps near the initial configuration $\lambda_2 = \lambda_3$, reflecting strong asymmetric perturbations between planets two and three at the onset of the integration.

By setting the initial angle $\lambda_1=0$°,  Eq. \ref{eq:3body_long} rearranges to express $\lambda_3$ as a function of $\lambda_2$:

\begin{equation}
\label{eq:stripes}
       \lambda_3 =  \frac{p+q}{q} \lambda_2 + \frac{\phi}{q},
\end{equation}

\noindent which represents the equation of a straight line of slope $(p+q)/q$ passing through the point ($\phi/q$, $\phi/q$).  If we assume that the stripes are the result of a three-body resonance, then Eq.\,\ref{eq:stripes} tells us that the number of stripes (counting from $\lambda_2 = 0$° to 360°) is given by the index of the resonance $p+q$ and the slope of the stripes by $(p+q)/q$. Fig.\,\ref{fig:stripes54135} shows 19 stripes, which means $p+q = 19$. With the slope being slightly less than two and 19 being an odd number, it is very likely that $q = p+1$. This therefore corresponds to the zeroth-order 3BR with $\alpha = 10/9$.

\begin{figure} 
\centering
\includegraphics[width=1.0\linewidth]{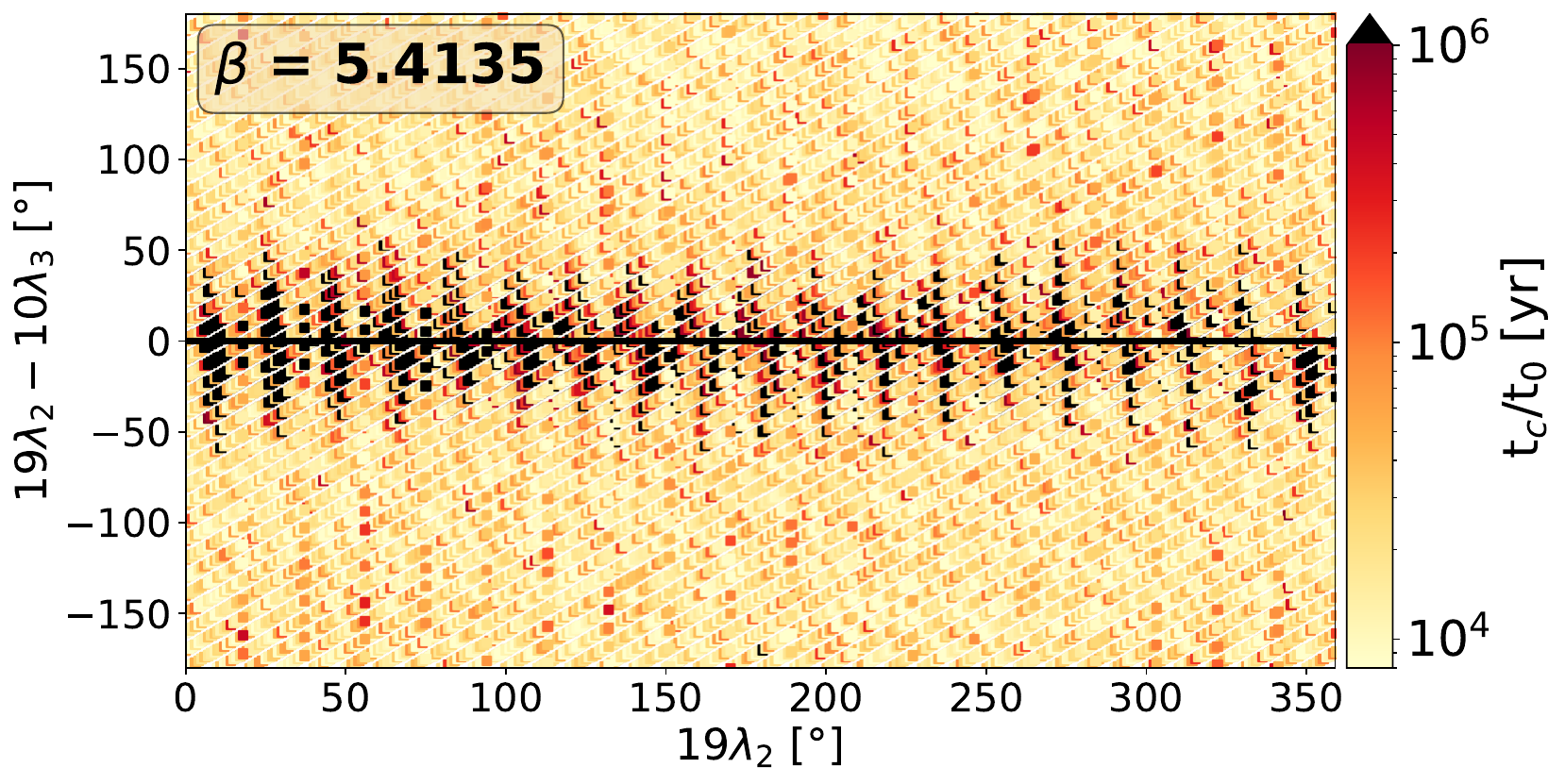}
\caption{ \label{fig:convergence54135} Lifetimes of systems $b541$ (dataset from Fig.\,\ref{fig:stripes54135}), distributed in the three-body resonance space with $\alpha=10/9$. Long-lived systems cluster around the resonant angle $\phi=0$°. Systems that survive for more than 10$^{6}$ yrs are shown in black.}
\end{figure}

\begin{figure*} 
\centering
\includegraphics[width=1.0\linewidth]{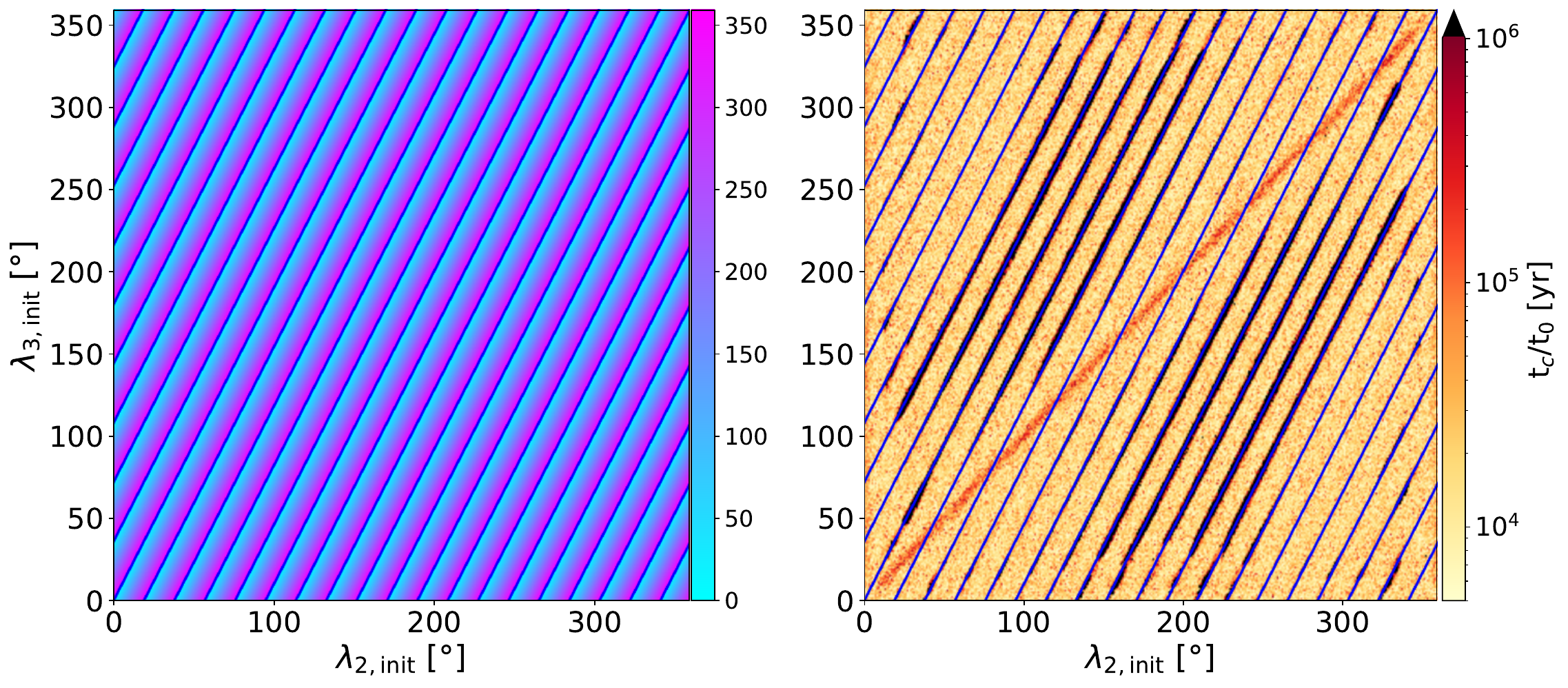}
\caption{Left: Resonant angle $\phi$ with $\alpha=10/9$ for all values $\lambda_2$ and $\lambda_3$. Dark blue stripes correspond to $\phi=0°$. Right: Same as Fig.\,\ref{fig:stripes54135}, but with the blue stripes overlaid. The angle $\phi=0$° overlaps with the stripes of long-lived systems. \label{fig:super_54135} }
\end{figure*}

We verified this assumption by examining whether stable systems clustered around a given angle $\phi$ for 3BR$\alpha10/9$. Figure\,\ref{fig:convergence54135} shows that all long-lived systems cluster around the resonant angle $- 19 \lambda_2 + 10 \lambda_3 = 0$°. Figure \,\ref{fig:super_54135} compares the map of $\phi$ for all angle values with the lifetime grid for $b541$. The long-lived stripes clearly overlap with the locations where $\phi =0°$, indicating that the dynamics of system $b541$ are driven by 3BR$\alpha\frac{10}{9}$.

\begin{figure} 
\centering
\includegraphics[width=1.0\linewidth]{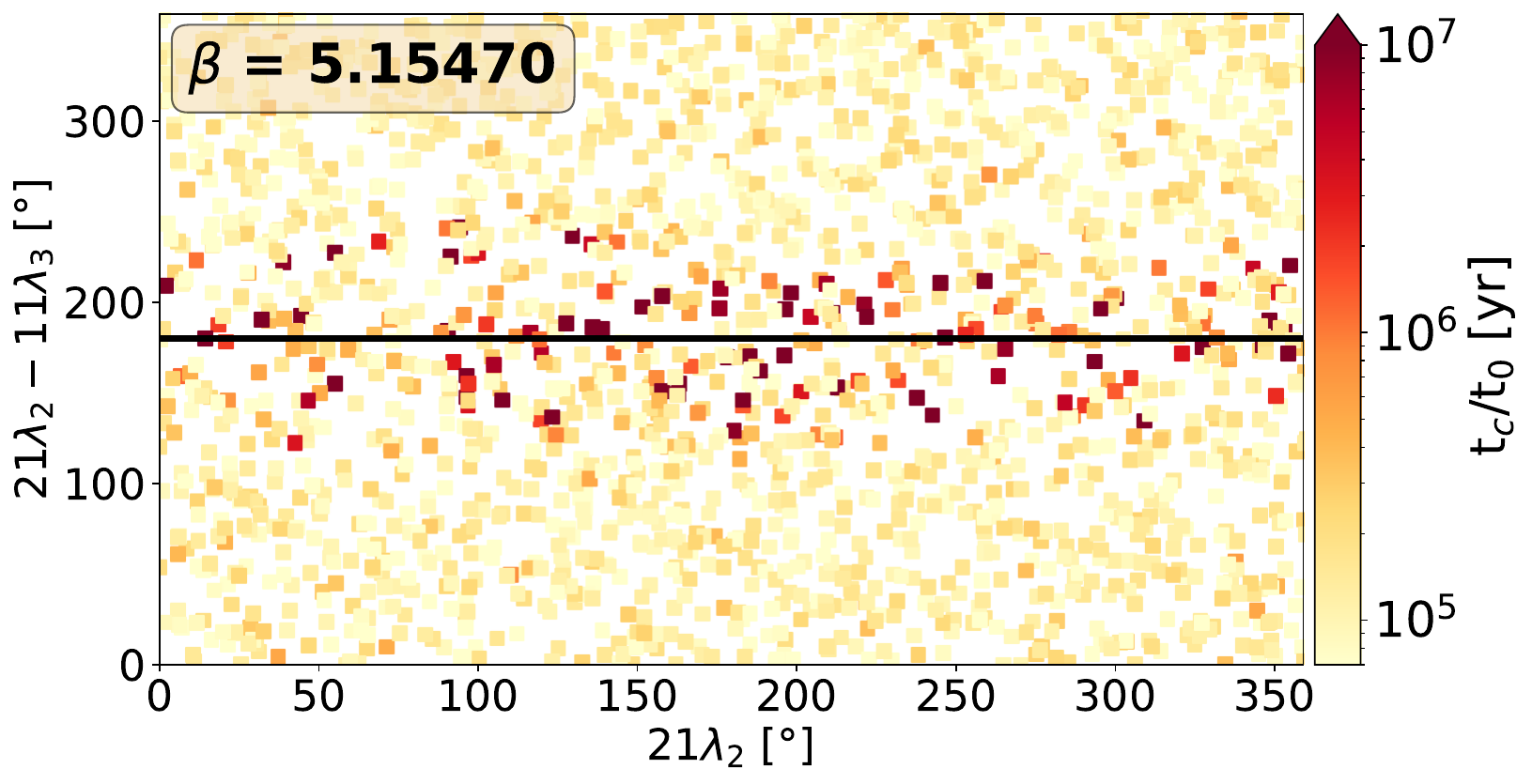}
\caption{Lifetimes of system $b515$ distributed in the three-body resonance space with $\alpha=11/10$. Long-lived systems cluster around the resonant angle $\phi=180$°.\label{fig:convergence51547} }
\end{figure}

We applied the same analysis to system $b515$. However, unlike $b541$, SPK4 is not in a dip (Fig.~\ref{fig:drawing}). Most systems in the vicinity of SPK4 have lifetimes about 2$\sigma$ above the exponential fit, corresponding to $t_c/t_0 \approx 10^{5}$ years at the location of $b515$, which would make the computation of a grid similar to Fig.\,\ref{fig:stripes54135} significantly more costly. We drew 2000 systems $b515$ from a random distribution of initial angles $\lambda_2$ and $\lambda_3$ and determined which pair of $p, q$ values produces the long-lived systems clustering around an angle $\phi$. Figure \ref{fig:convergence51547} shows that the $b515$ systems cluster around $\phi=180$° for the pair $\alpha=11/10$.

\begin{figure} 
\centering
\includegraphics[width=1.0\linewidth]{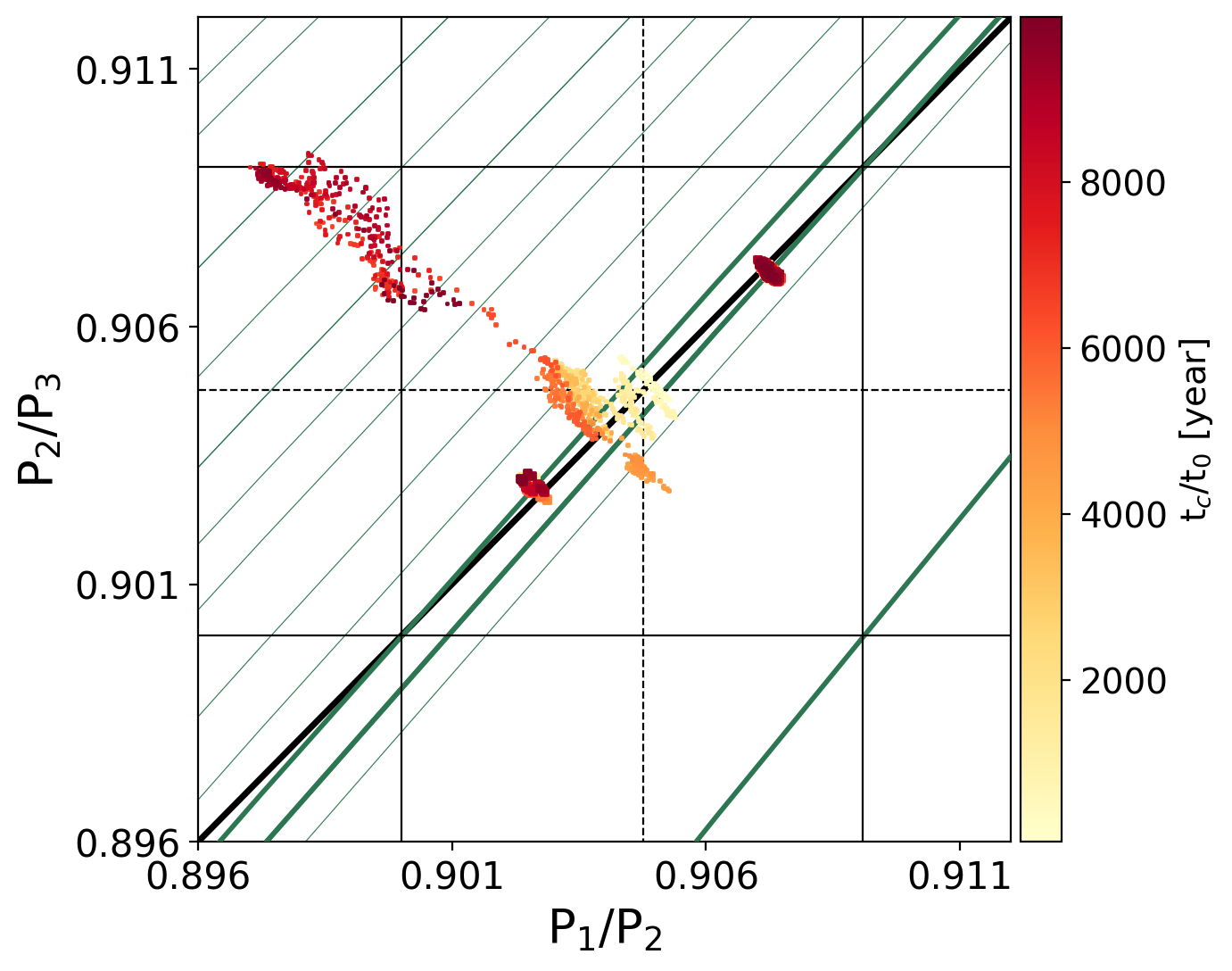}
\caption{Trajectory over the first 10,000 years, with a 20-year moving average, in the period-ratio plane for three selected systems: one in SPK4 ($\beta$ = 5.1547), one in SPK5 ($\beta$ = 5.4135), and one midway in the vicinity of a second-order 2BR intersection ($\beta$ = 5.2813), sharing the same initial longitudes as the system in SPK4. Vertical and horizontal solid and dashed black lines correspond to first- and second-order two-body MMRs, respectively, between adjacent planet pairs. The solid diagonal black line in the upper right corner indicates the first-order MMRs between the inner and outer planets. The oblique green lines correspond to the zeroth-order three-planet MMRs. Only the 3BRs with $p+q < 24$ are shown. The three bold green curves correspond to 3BRs with, from left to right, $\alpha=11/10$, $\alpha=10/9$, and $\alpha=1$. \label{fig:pratio_spk45} }
\end{figure}

Figure\,\ref{fig:pratio_spk45} shows the trajectories of systems $b515$ and $b541$. In this region, the 3BR network is very dense, and the width of the two resonances overlaps. No noticeable diffusion occurs on the plane during the whole integration (up to 10$^{10}$ yrs) and both systems remain away from the 2BR network (of first- and second-order 2BRs). All spike systems observed in \citetalias{Lissauer&Gavino_2021} are also located away from 2BRs, whereas dips occur in the vicinity of 2BRs. For comparison, we show the trajectory of a system midway between $b515$ and $b541$. This system begins to diffuse perpendicular to the 3BR network (as expected by Chirikov diffusion) within the first few thousand orbits.

\subsection{Spikes SPK1, SPK2, and SPK3}\label{sec:spikes123}

Figure \,\ref{fig:anomalous} shows an excess of long-lived systems with the outer pair of planets initially near conjunction. This result should be approached with caution because integrations that start near conjunctions begin with a highly asymmetric transfer of energy and angular momentum from the planet closer to the star (here, the middle planet) to the outer planet. This transfer produces a wider separation between the orbits of these two planets and, if the innermost planet starts far from alignment, reduces the separation of the inner pair by roughly half as much, effectively moving the system away from equal initial period ratios. The system is thus ``pushed'' toward the lower right in the period ratio plane.

\begin{figure} 
\centering
\includegraphics[width=1.0\linewidth]{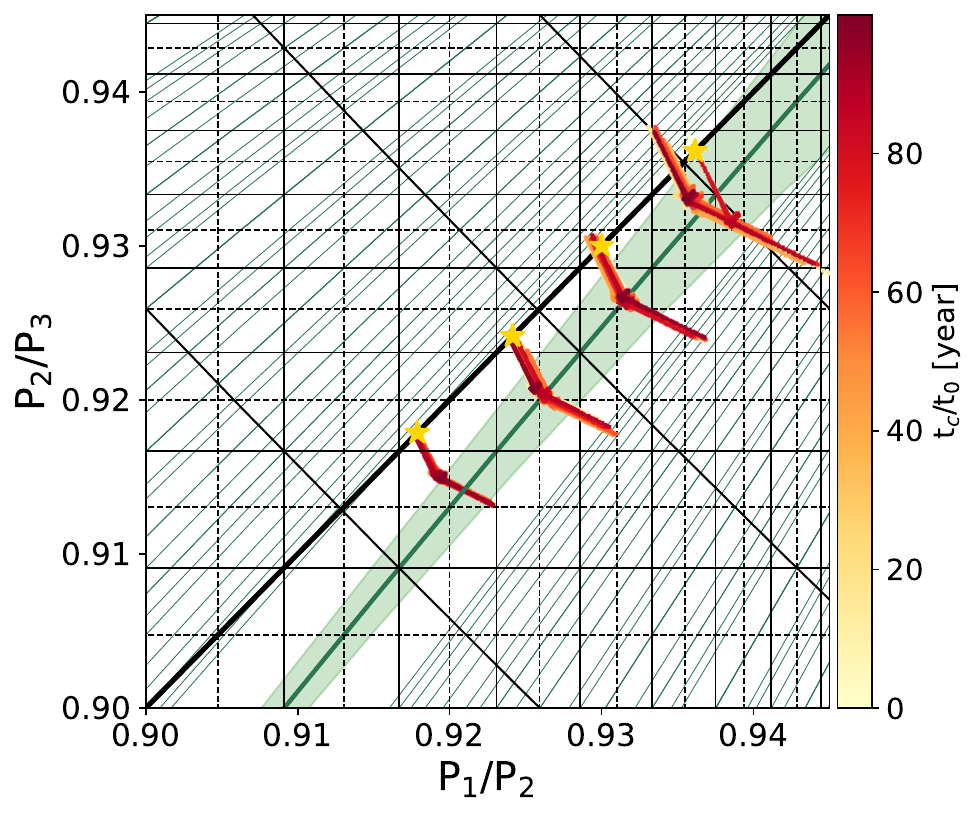}
\caption{Trajectories over the first 100 years for selected systems in SPK1 ($\beta$ = 3.82086, $\lambda_2 = 166.273784$°, $\lambda_3 = 165.110014$°), SPK2 ($\beta$ = 4.15423, $\lambda_2 = 179.01980$°, $\lambda_3 = 178.62$°), and SPK3 ($\beta$ = 4.51276, $\lambda_2 = 211.95$°, $\lambda_3 = 212.22$°). For comparison, we also show the longest-lived closely spaced system in the S2 set ($\beta$ = 3.4744, $\lambda_2 = 207.16381$°, $\lambda_3 = 208.66160$°), located near the upper right corner. Positions are plotted every 0.1 years; gold stars mark the initial period ratios. Solid and dashed black lines indicate first- and second-order two-body MMRs, respectively. The bold black diagonal line shows equally spaced orbits, and oblique green lines indicate the zeroth-order three-body MMRs for $p+q < 24$. The shaded green area indicates the width (Eq.~55 of \citetalias{Petit+etal_2020}) of the resonance with $\alpha = 1$. \label{fig:res_width} }
\end{figure}

We selected one representative system from each of SPK1, SPK2, and SPK3. For SPK1, we used the last survivor ($t_c/t_0 \geq 10^{10}$ years) from S1 with $\beta = 3.82086$ (hereafter $b382$). For SPK2 and SPK3, we selected the longest-lived systems from the standard random set, with $\beta = 4.15423$ and $\beta = 4.51276$ (hereafter $b415$ and $b451$, respectively). All three systems have outer pairs initially near conjunction. Figure \ref{fig:res_width} shows their trajectories over the first 100 years, revealing characteristic ``wings'' caused by conjunctions: inner pair conjunctions produce lower-right wings, while outer pair conjunctions produce upper-left wings. The integrator interprets the initial configuration as the second half of a conjunction\footnote{For this reason, integrations are usually initialized far from conjunctions (e.g., \citealt{Smith+Lissauer_2009}).}. For low eccentricities, the conjunctions are normally symmetric and the system returns to its initial position. When only half of the interaction is captured, the energy exchange is not canceled , and the system goes in a wrong direction. The system is displaced toward the vicinity of the isolated 3BR$\alpha1$, which explains its anomalous stability. Studies with random initial longitudes show similar effects, although additional planets often destabilize such configurations in higher-multiplicity systems.

\begin{figure} 
\centering
\includegraphics[width=1.0\linewidth]{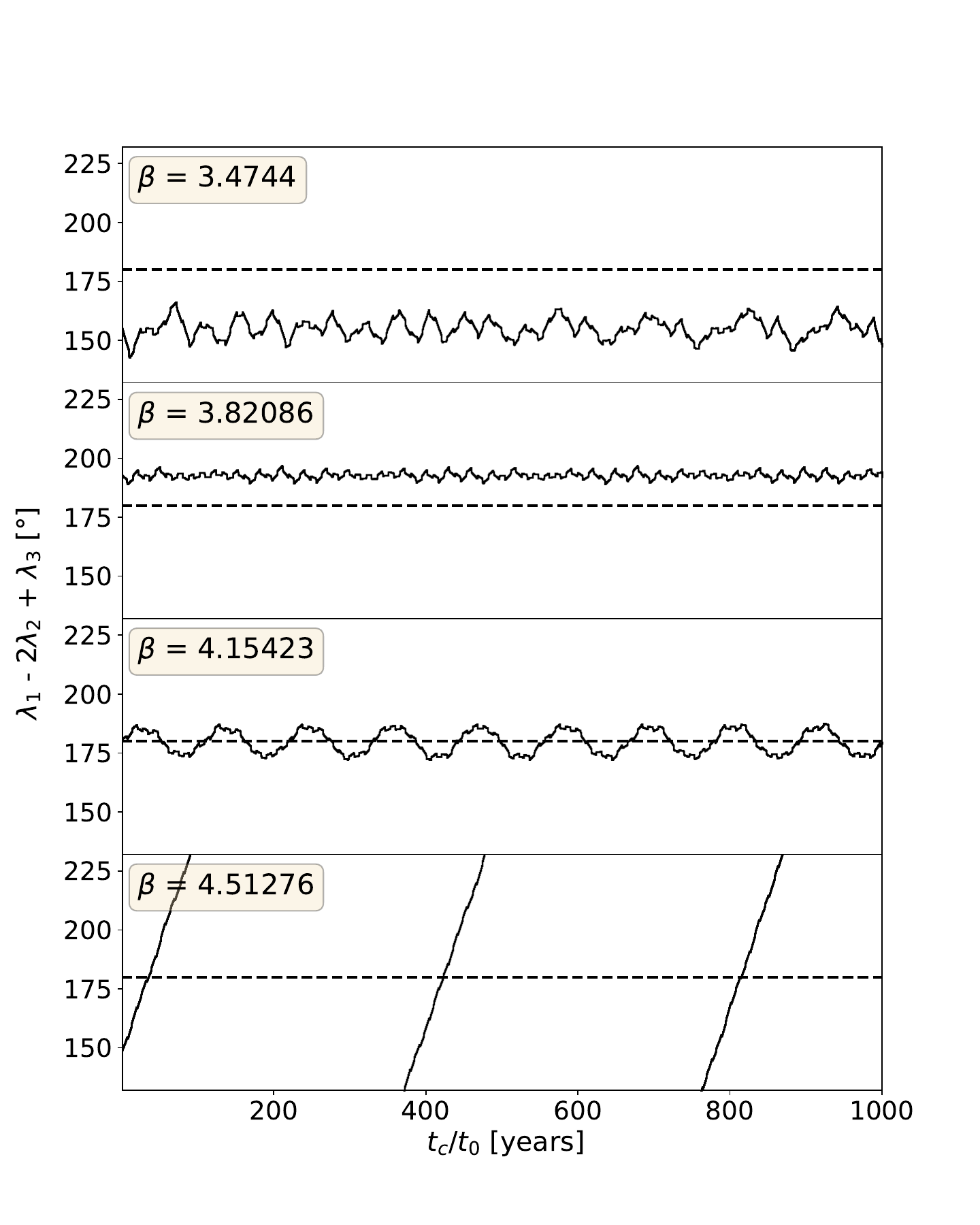}
\caption{Evolution of three-body resonant angles over the first 1000 orbits of four systems. Top panel: Relatively long-lived, extremely closely spaced, system in S2 set ($\beta$ = 3.4744, $\lambda_2 = 207.16381$°, $\lambda_3 = 208.66160$°). Second panel: System in SPK1 ($\beta$ = 3.82086, $\lambda_2 = 166.273784$°, $\lambda_3 = 165.110014$°). Third panel: System in SPK2 ($\beta$ = 4.15423, $\lambda_2 = 179.02$°, $\lambda_3 = 178.62$°). Bottom panel: SPK3 ($\beta$ = 4.51276, $\lambda_2 = 211.95$, $\lambda_3 = 212.22$°).  \label{fig:resangle1} }
\end{figure}

To illustrate the dynamics with respect to 3BR$\alpha1$, Fig. \ref{fig:resangle1} shows the evolution of the resonant angle over the first 1000 years. The system $b415$ librates around the resonant angle ($-2 \lambda_2 + \lambda_3 = 180$°), whereas the system $b451$ circulates, which is expected because the system settled just outside 3BR$\alpha 1$. The resonant angle for $b382$ librates around $-2 \lambda_2 + \lambda_3 = 192.57$° with a very small amplitude and without a clear periodic pattern. None of these systems experiences a triplet close encounter during the first 1000 years. The system $b382$ does not experience any triplet close encounters over the total integration time of 10$^{10}$ yrs and remains effectively locked in this 3BR throughout the integration (note that not experiencing triplet close encounters is not a requirement for stability). The systems $b382$ and $b415$ are also near the intersection of two second-order 2BRs (27:29 and 25:27 for $b382$, 25:27 and 23:25 for $b415$), but none of these two-body resonant angles librates (not shown), indicating that both systems are locked in a true 3BR \citep{Papaloizou+etal_2015, Gozd+etal_2016}. The system with $\beta = 3.4744$ in the S2 set also librates around $-2\lambda_2 + \lambda_3 \approx 150°$. Unlike SPK4 and SPK5, the systems discussed here (SPK1, SPK2, and SPK3) remain near 2BR intersections (see Appendix \ref{app:2br}). 

In Fig.\,\ref{fig:drawing} SPK3 appears split into two peaks separated by the 23:25 line. This splitting likely results from libration of the three-body resonant angle, which allows interaction with the same resonance at slightly different period ratios. A similar split occurs in SPK1 (Fig.,\ref{fig:special1}) and reflects the resonance width rather than distinct resonances.

\begin{figure} 
\centering
\includegraphics[width=1.0\linewidth]{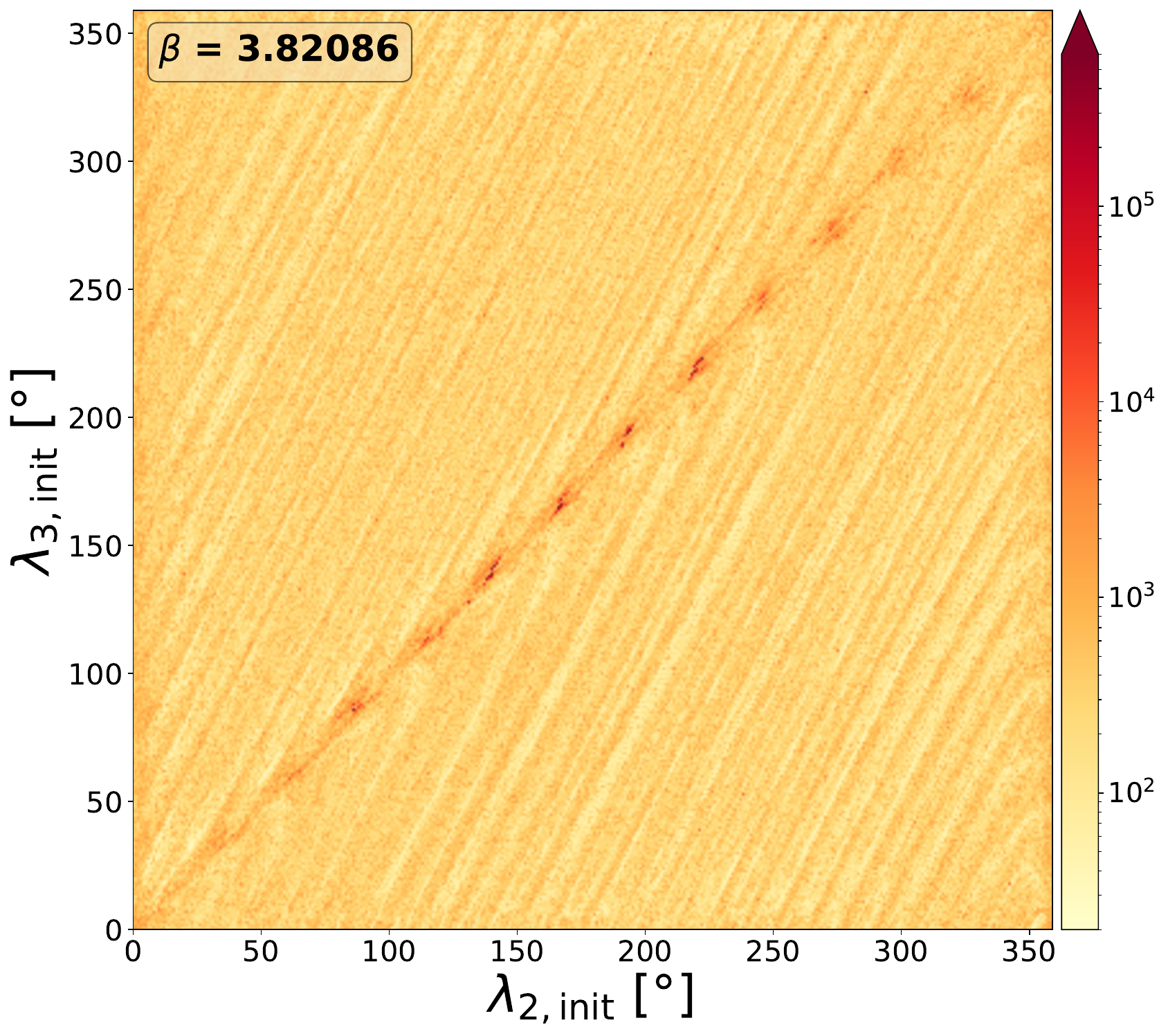}
\caption{System lifetime as a function of the initial longitudes of the middle ($\lambda_2$) and outer ($\lambda_3$) planets. The system has an initial $\beta = 3.82086$, selected from the spike in Specific Set S1, with a lifetime of 10$^{10}$ years. The angular resolution is 1°, corresponding to a total of 360$\times$360 integrations. \label{fig:stripes382086} }
\end{figure}

Figure\,\ref{fig:stripes382086} shows, for the case of $b382$, the results of an analysis similar to that in Sect.\,\ref{sec:spikes45} and  Fig.\,\ref{fig:stripes54135}. Interpreting the result is not as straightforward as in the case of $b541$. The surface shows 13 successive small long-lived islands along the diagonal $\lambda_2 = \lambda_3$ and a succession of faint stripes throughout the surface. Fourteen stripes should pass through the long-lived islands. If we follow the same analysis as in Sect.\,\ref{sec:spikes45}, we find $p+q=14$, such that $\alpha=4/3$ or $\alpha=1$. However, Figs.\,\ref{fig:res_width} and \ref{fig:resangle1} show that the dynamics is at least partially regulated by 3BR$\alpha1$, and therefore $\alpha=4/3$ can be excluded. Since only the initial configurations close-to-conjunction can push the system toward 3BR$\alpha1$, long-lived resonant systems are necessarily near the line $\lambda_2 = \lambda_3$. The small long-lived islands in Fig.\,\ref{fig:stripes382086} correspond to the portions of the 14 stripes that lie close to the $\lambda_2 = \lambda_3$ diagonal.  Appendix \ref{app:3br} shows this for a system that starts directly on 3BR$\alpha1$.

\subsection{The wide region between spikes}\label{sec:gap}

As mentioned in Sect.\,\ref{sec:random:tc}, there is a wide region without spikes between SPK3 and SPK4. In Sect.\,\ref{sec:spikes123}, we show that the systems are ``pushed'' toward 
3BR$\alpha1$, where they remain locked in a stable configuration. Furthermore, Sect.\,\ref{sec:3br} shows that 3BR$\alpha1$ is special, as it is the most isolated resonance within the 3BR network. In Fig.\,\ref{fig:res_width}, the green shaded area represents the width of the resonance, as defined by \citetalias{Petit+etal_2020} (their Eq.\,55). Because 3BR$\alpha1$ is located on the lower-right side of the diagonal, systems starting on the main diagonal with the outer pair initially close to conjunction can be captured by this resonance (see Sect.\,\ref{sec:spikes123}). 

The distance in the period-ratio plane to which a system can be ``pushed'' by a conjunction depends on the initial orbital spacing. If the initial orbital spacing is too large, the energy exchange during the two-planet conjunction cannot move the systems into the vicinity of the resonance. Figure~\ref{fig:res_width} shows that the system in SPK3 barely reaches the resonance width, suggesting that no other spikes due to this resonance can occur for larger $\beta$ values. In this sense, SPK3 is a transition spike. The subsequent spikes (SPK4, SPK5, and beyond) can only result from capture by a 3BR that intersects the main diagonal.

\section{Unequally spaced initial orbits}\label{sec:nonequally}

Sect.\,\ref{sec:spikes123} shows that the long-lived systems in the inner spikes possess nonuniform proper elements. However, this does not invalidate the fact that these systems are stable; they simply do not belong to the ``proper'' main diagonal. This motivates further numerical investigations, which require integrations of systems outside the main diagonal, i.e., systems with non-evenly spaced orbits. This section focuses on 3BR$\alpha1$.

The stability of systems locked in 3BR$\alpha1$ is primarily governed by their position relative to the 2BR network. As shown in Sect.\,\ref{sec:3br}, 3BR$\alpha1$ systematically passes through intersections of same-order 2BRs, so both planet pairs share the same dynamical behavior. In first-order 2BRs, successive conjunctions occur at nearly the same longitude in the inertial frame (modulo apse precession), whereas in second-order 2BRs they occur near opposite longitudes \citep{Murray+Dermott_1999}. This can lead to cancellation effects, especially for higher-order resonances and eccentric orbits \citep[e.g.,][]{Peale_1976, Tamayo+Hadden_2025}. As a result, a system locked in 3BR$\alpha1$ at the intersection of two $k$th-order 2BRs returns to the same configuration every $k \times P_\mathrm{3br}$, and conjunction cancellation can enhance stability for favorable longitudes (beyond the scope of this paper).

This framework explains the lifetime pattern near 3BR$\alpha1$. First-order 2BRs tend to excite eccentricities, increasing period ratios and producing lifetime dips just narrow of these resonances \citep[][]{Lissauer+Espresate_1998,Espresate+Lissauer_2001,Obertas+etal_2017,Lissauer&Gavino_2021,Lammers+etal_2024}. Systems near, or slightly outside, intersections of first-order 2BRs diffuse parallel to the main diagonal (and 3BR$\alpha1$ for period ratios near unity), remaining protected from Chirikov diffusion and forming stable islands. In contrast, second-order 2BRs are weak at zero eccentricity, allowing systems to remain in 3BR$\alpha1$ and achieve secular stability for favorable longitudes. Stable islands are therefore expected directly at the intersections of second-order 2BRs.

We used Eq.\,\ref{eq:beta} to define two different orbital separations, one for the inner pair $\beta_{12}$ and another for the outer pair $\beta_{23}$, and derived the initial orbital elements accordingly. To limit the computation time, we restricted the integrations to a small surface of the period-ratio plane ranging from the Hill stability limit to $\beta_{12} + \beta_{23} = 10$, which corresponds to the shape of a triangle in the $(\beta_{12},~\beta_{23})$ plane, and we stopped the integration if the system remained stable for at least 10$^7$ years. We selected a single, widely spaced set of initial longitudes, $\lambda_2 = 222.49$° and $\lambda_3 = 84.98$° previously used in \citet{Smith+Lissauer_2009}, which form the ``SL09'' set in \citetalias{Lissauer&Gavino_2021}. All integrations start far from any planet conjunctions, and this set ensures that 3BR$\alpha1$ has a value $\phi = 0$°. 

\begin{figure} 
\centering
\includegraphics[width=1.0\linewidth]{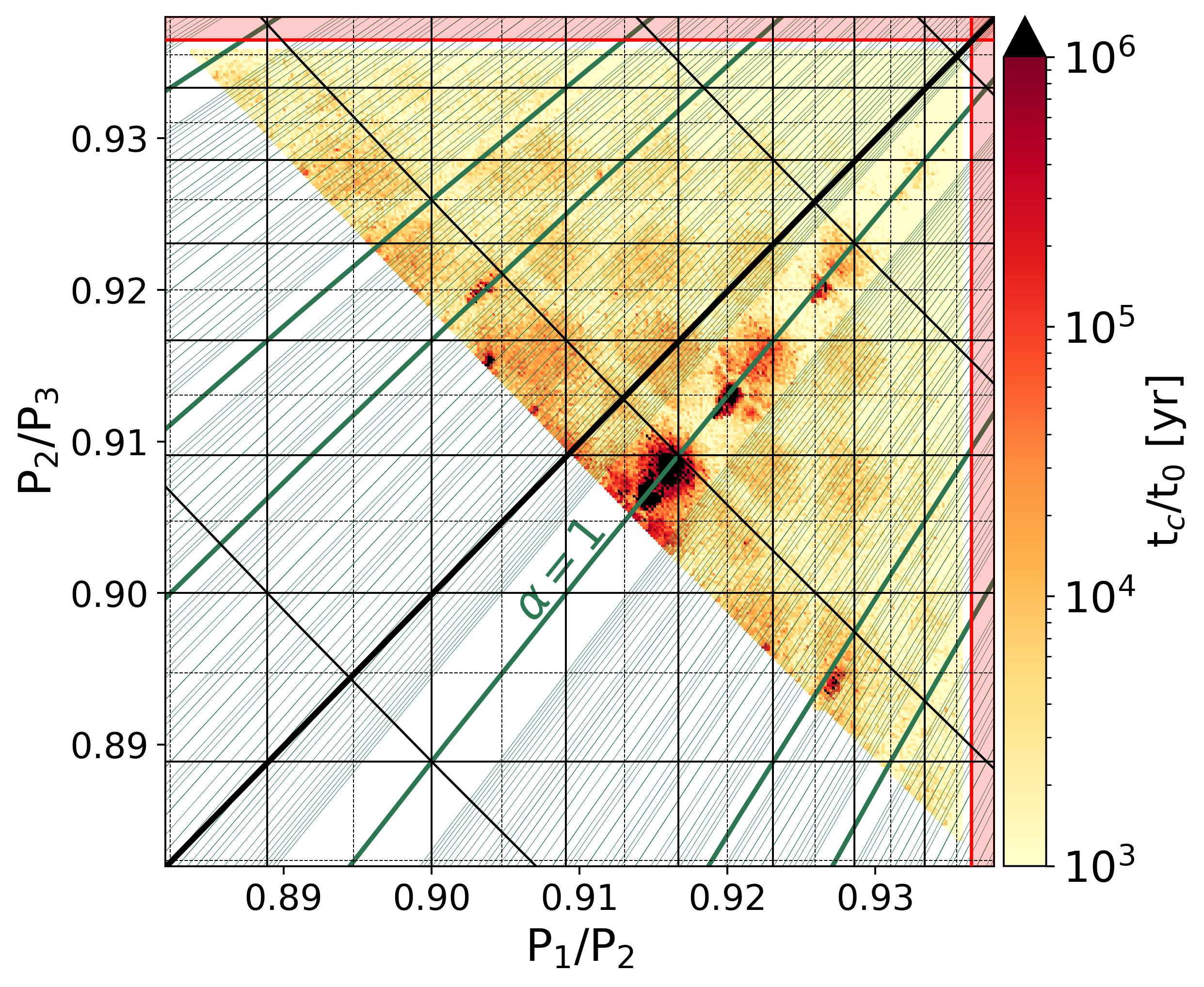}
\caption{Lifetime (color map) of systems with unevenly spaced orbits shown in the period-ratio plane. Integrations are performed over a surface ranging from the Hill stability limit to $\beta_{12} + \beta_{23} = 10$. The resolution in $P_j/P_{j+1}$ is $\sim 1.90\times10^{-4}$ (or 0.01 in $\beta$). The initial longitudes are $\lambda_2 = 222.49$° and $\lambda_3 = 84.98$°, the same as in \citet{Smith+Lissauer_2009} and \citetalias{Lissauer&Gavino_2021}, and correspond to a resonant angle $-2\lambda_2 + \lambda_3 \approx 0°$. The diagonal black line represents systems with evenly spaced orbits, and the oblique green lines indicate the network of three-body commensurability. The bold green lines indicate resonances in less dense regions, including 3BR$\alpha1$ and 3BR$\alpha3/2$ in the upper left. In the vicinity of 3BR$\alpha1$, 276 systems survived for more than 10$^6$ years, including 60 that survived for more than 10$^7$ years. Elsewhere (other isolated islands), 19 systems survived for more than 10$^6$ years and 0 for more than 10$^7$ years. Systems with a lifetime greater than 10$^6$ years are shown in black.\label{fig:gridlifetime} }
\end{figure}

Fig.\,\ref{fig:gridlifetime} shows the numerical results, which displays a grid of relatively stable islands with a typical lifetime between 5000 and 10,000 years. This grid of quasi-stable islands results directly from the 2BR network, as suggested above. The additional presence of anomalously long-lived islands is immediately apparent. These islands, which also follow the overall 2BR grid pattern, are due to the 3BR network. Long-lived islands systematically lie near well-isolated 3BRs. In particular, 3BR$\alpha1$ hosts a predicted chain of very long-lived islands. These comprise small long-lived islands lying on top of the intersection of second-order 2BRs and larger, long-lived islands directly outside the intersection of first-order 2BRs. These islands appear confined and shaped by the 3BR network; this also applies to other long-lived islands on top of other 3BRs. This confinement indicates that the locations where the systems can remain stable are regulated by 2BRs, while the protection offered by well-isolated 3BRs allows the systems to become anomalously long-lived, up to 10$^{10}$ years (see Fig.~\ref{fig:special1}), even in regions very close to the Hill stability limit and well within the overlap stability limit of 3BRs.

\section{Discussion}

Figure\,\ref{fig:gridlifetime} illustrates why extremely closely spaced three-planet systems are more likely to remain stable near isolated 3BRs. Here, we briefly discuss tightly spaced {\it Kepler} systems with at least three planets and their position on the 3BR network, particularly near isolated resonances. As discussed in Sect.\,\ref{sec:3br}, these resonances have more overlapping components, resulting in a high density of 2BR crossings and increased isolation from the surrounding 3BR network.

\begin{figure} 
\centering
\includegraphics[width=1.0\linewidth]{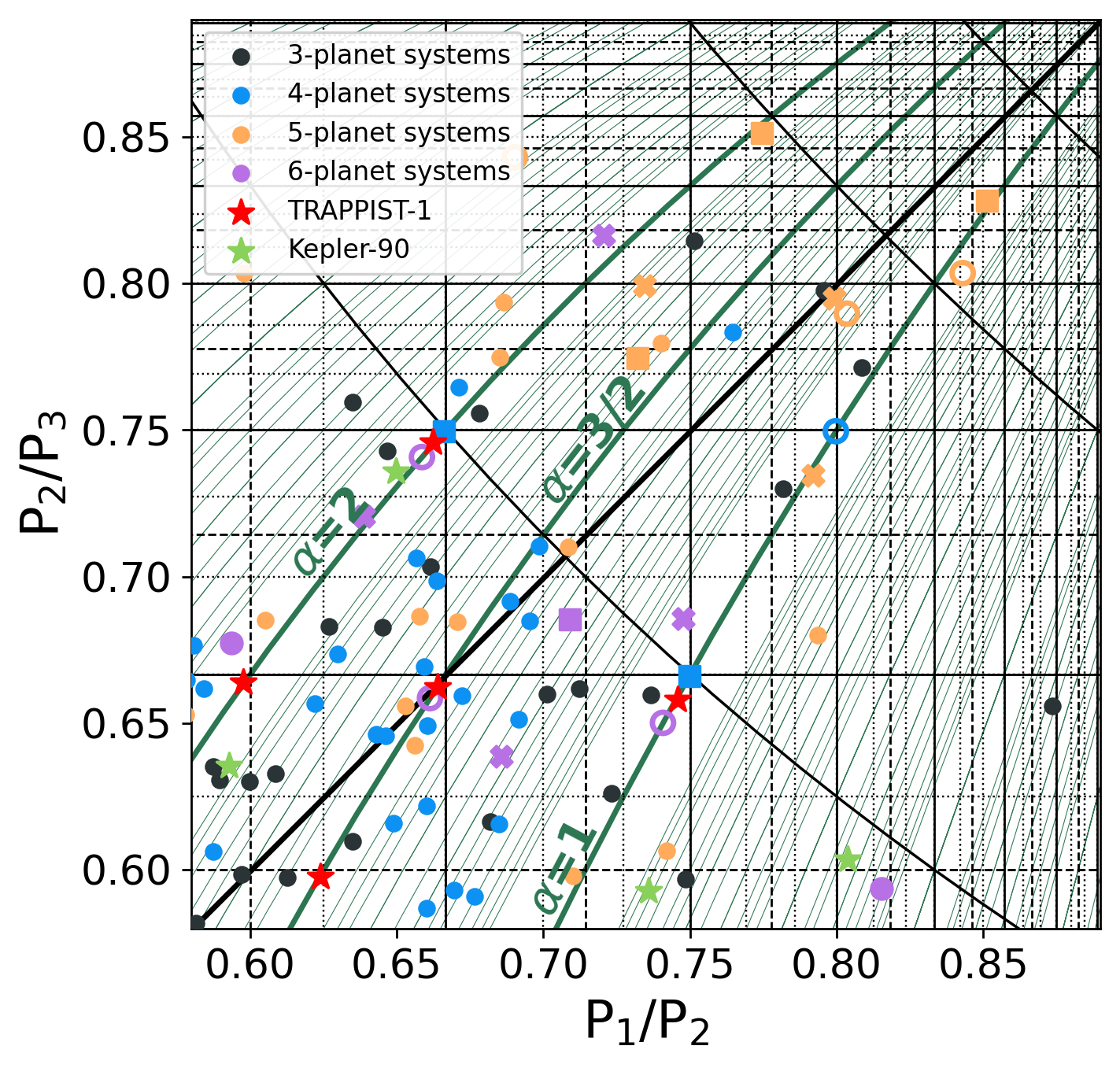}
\caption{Period-ratio plane showing adjacent triplets in {\it Kepler} systems within $0.58 < P_j/P_{j+1} < 0.90$ \citep{Lissauer+etal_2024}. Black markers denote triplets in three-planet systems. Blue markers show triplets in four-planet systems (up to two per system), including Kepler-223 (squares) and the resonant triplet in Kepler-60 (open circle). Yellow markers show triplets in five-planet systems (up to three per system), including Kepler-1371 (open circles), Kepler-1542 (squares), and Kepler-444 (crosses). Purple markers show triplets in six-planet systems (up to four per system), including Kepler-80 (open circles), Kepler-11 (squares), and Kepler-102 (crosses). Light green stars denote triplets in the eight-planet system Kepler-90, and red stars denote triplets in TRAPPIST-1. Solid, dashed, and dotted black lines indicate first-, second-, and third-order two-body mean-motion resonances. Oblique green lines show zeroth-order three-body resonances; bold lines mark the most isolated resonances with $\alpha=1$, $3/2$, and $2$.\label{fig:gridkepler} }
\end{figure}

Figure\,\ref{fig:gridkepler} shows triplets of tightly spaced {\it Kepler} planets on the period-ratio plane. Among them are notable tightly spaced chains of resonances and planet triplets known to be captured in 3BRs. The Kepler-60 system \citep{Steffen+etal_2013, Gozd+etal_2016, Jontof+etal_2016} hosts a triplet that lies in 3BR$\alpha1$ and near the intersection of two first-order 2BRs \citep[4:5 and 3:4; e.g.][]{Gozd+etal_2016}. Kepler-223 has four confirmed planets \citep{Lissauer+etal_2014,Rowe+etal_2014, Mills+etal_2016} and therefore two consecutive triplets. Both triplets lie on the main 3BRs ($\alpha1$ for the inner triplet and $\alpha2$ for the outer triplet) and each adjacent pair is also on top of a first-order 2BR (3:4, 2:3, 3:4). 

Kepler-11 \citep{Lissauer+etal_2011a}, Kepler-80 \citep{Lissauer+etal_2011b, Shallue+_2018, MacDonald+etal_2021}, Kepler-102, and Kepler-90 are the only {\it Kepler} systems with six planets in which at least two triplets are represented in Fig.\,\ref{fig:gridkepler}. The triplet positions on the period-ratio plane differ fundamentally between these systems. Kepler-80 is one of the most notable cases of resonant chains \citep{MacDonald+etal_2016, MacDonald+etal_2021}, where all triplets among the five planets in the chain lie near the main isolated 3BRs: 3BR$\alpha$1, 3BR$\alpha$2, and 3BR$\alpha$3/2 for the innermost, middle, and outermost triplets, respectively. The triplets of Kepler-11 triplets lie far from the main isolated 3BRs. Similarly, both Kepler-90 triplets lie far from these isolated 3BRs. Kepler-102 is the only six-planet system that contains four visible triplets, i.e., all adjacent pairs have period ratios $> 0.58$. Two of them lie near 3BR$\alpha$2, one near the 3BR$\alpha$1, and another away from any 2BRs and isolated 3BRs. These systems have very closely packed adjacent planets with similar period ratios, but exhibit very different architectures, which is commonly attributed to different formation mechanisms \citep[e.g.,][]{Hands+etal_2014}. Most {\it Kepler} planets have a mass ratio to their star that is a factor of $\sim 2 - 8$ times larger than the values we simulated in this paper, but some {\it Kepler} planets have substantially smaller masses. Figure\,\ref{fig:gridkepler} also shows the position of the adjacent triplets in the TRAPPIST-1 system \citep{Guillon+2017, Luger+etal_2017, Agol+etal_2021}. As with the {\it Kepler} resonant triplets, all adjacent triplets in TRAPPIST-1 lie on the main isolated 3BRs and in the vicinity of first-, second-, and third-order 2BR intersections. 

More globally, \citet{Cerioni+etal_2022} used a statistical analysis to reveal a correlation between the observed distribution of planet triplets and the main isolated 3BRs. Most of the very tightly packed resonance chains lie near the main isolated 3BRs, suggesting that these resonances are more fundamental than the 2BR network in this area of the period-ratio plane. Fig.\,\ref{fig:gridlifetime} directly illustrates that the stability of very tightly packed resonant systems cannot be understood without addressing the interplay between the 2BR and the 3BR networks, a mechanism that should be included as a feature in stability prediction models \citep[e.g.,][]{tamayo+etal_2016, Thadhani+etal_2025}. Extremely tightly packed resonant chains are expected to lie on these isolated 3BRs because they are favorable locations on the period-ratio plane protected from Chirikov diffusion (for period ratios close to one, a system can diffuse roughly parallel to the 3BR network only if it is captured by these resonances).

\section{Conclusions}

We used numerical simulations to investigate the dynamics of extremely compact three-planet systems ($P_{j}/P_{j+1} \gtrapprox 0.90$ or $\beta \lessapprox 5.6$). We find a small set of initial conditions that allow three-planet systems more tightly packed than any observed systems to remain dynamically stable for billions of orbits. Building on previous analytical studies \citep[e.g.,][]{Quillen_2011, Petit+etal_2020} and observed systems, we present a qualitative analysis of the role of three-body resonances and the relationship between their initial longitudes. 

First, we argue that the most isolated three-body resonances, specifically 3BR$\alpha2/3$, 3BR$\alpha1$, 3BR$\alpha3/2$, and 3BR$\alpha2$, are favorable locations for stabilizing extremely tightly packed three-planet systems. Second, we show that overall stability depends on the interplay between these isolated three-body resonances and the surrounding two-body resonance network. We summarize the results as follows:

\begin{itemize}
\item In numerical simulations, we observe anomalously stable systems with extremely small period ratios that remain stable over very long times. This stability correlates with the initial longitudes. 

\item The dependence of stability on initial longitudes reflects capture into a three-body resonance. Extremely compact three-planet systems remain stable only if their dynamics are regulated by three-body resonances, in particular a specific subset of resonances.

\item Some three-body resonances are more isolated than others in the network. We show that the degree of isolation of a three-body resonance is regulated by the two-body resonance network; three-body resonances cross intersections of two-body resonances, and the degree of isolation is proportional to the density of intersection crossings, which depends on the ratio $\alpha \equiv q/p$. The most isolated three-body resonances have $\alpha = 1/2; 2/3; 1; 3/2; 2$. We leave this analysis for future work. 

\item This isolation is the main protection against chaotic diffusion for extremely tightly packed three-planet systems, particularly those that lie inside the three-body resonance overlap limit and have $P_{j}/P_{j+1} > 0.90$. If observed, extremely tightly packed stable planet triplets, similar in size to or larger than Earth, are expected to lie near this subset of the three-body resonance network.

\end{itemize}

\begin{acknowledgements}
We thank Antoine Petit, Tjarda Boekholt, and Dan Sirbu for crucial discussions. S.G. acknowledges support from the ERC Synergy Grant “ECOGAL” (project ID 855130) and the Independent Research Fund Denmark (grant No. 0135-00123B).
\end{acknowledgements}

\bibliography{ref}

@ARTICLE{Peale_1976,
       author = {{Peale}, S.~J.},
        title = "{Orbital resonance in the solar system.}",
      journal = {\araa},
         year = 1976,
        month = jan,
       volume = {14},
        pages = {215-246},
          doi = {10.1146/annurev.aa.14.090176.001243},
       adsurl = {https://ui.adsabs.harvard.edu/abs/1976ARA&A..14..215P}
}

@ARTICLE{Hill_1878a,
       author = {{Hill}, G.~W.},
        title = "{Researches in the Lunar Theory.}",
      journal = {American Journal of Mathematics},
         year = 1878,
       volume = {1},
        pages = {5-26}
}

@ARTICLE{Chirikov_1979,
       author = {{Chirikov}, Boris V.},
        title = "{A universal instability of many-dimensional oscillator systems}",
      journal = {\physrep},
         year = 1979,
        month = may,
       volume = {52},
       number = {5},
        pages = {263-379},
          doi = {10.1016/0370-1573(79)90023-1},
       adsurl = {https://ui.adsabs.harvard.edu/abs/1979PhR....52..263C}
}

@ARTICLE{Wisdom_1980,
       author = {{Wisdom}, J.},
        title = "{The resonance overlap criterion and the onset of stochastic behavior in the restricted three-body problem}",
      journal = {\aj},
         year = 1980,
        month = aug,
       volume = {85},
        pages = {1122-1133},
          doi = {10.1086/112778},
       adsurl = {https://ui.adsabs.harvard.edu/abs/1980AJ.....85.1122W}
}

@INPROCEEDINGS{Aksnes_1988,
       author = {{Aksnes}, K.},
        title = "{General formulas for three-body resonances.}",
    booktitle = {Long-term Dynamical Behaviour of Natural and Artificial N-body Systems},
         year = 1988,
       editor = {{Roy}, Archie D.},
        month = jan,
        pages = {125-139},
       adsurl = {https://ui.adsabs.harvard.edu/abs/1988ltdb.conf..125A}
}

@ARTICLE{Gladman_1993,
       author = {{Gladman}, Brett},
        title = "{Dynamics of Systems of Two Close Planets}",
      journal = {\icarus},
         year = 1993,
        month = nov,
       volume = {106},
       number = {1},
        pages = {247-263},
          doi = {10.1006/icar.1993.1169},
       adsurl = {https://ui.adsabs.harvard.edu/abs/1993Icar..106..247G}
}

@ARTICLE{Chambers+etal_1996,
       author = {{Chambers}, J.~E. and {Wetherill}, G.~W. and {Boss}, A.~P.},
        title = "{The Stability of Multi-Planet Systems}",
      journal = {\icarus},
         year = 1996,
        month = feb,
       volume = {119},
       number = {2},
        pages = {261-268},
          doi = {10.1006/icar.1996.0019},
       adsurl = {https://ui.adsabs.harvard.edu/abs/1996Icar..119..261C}
}

@ARTICLE{Lissauer+Espresate_1998,
       author = {{Lissauer}, Jack J. and {Espresate}, Julia},
        title = "{Resonant Satellite Torques on Low Optical Depth Particulate Disks. I. Analytic Development}",
      journal = {\icarus},
         year = 1998,
        month = jul,
       volume = {134},
       number = {1},
        pages = {155-162},
          doi = {10.1006/icar.1998.5922},
       adsurl = {https://ui.adsabs.harvard.edu/abs/1998Icar..134..155L}
}

@ARTICLE{Espresate+Lissauer_2001,
       author = {{Espresate}, Julia and {Lissauer}, Jack J.},
        title = "{Resonant Satellite Torques on Low Optical Depth Particulate Disks. II. Numerical Simulations}",
      journal = {\icarus},
         year = 2001,
        month = jul,
       volume = {152},
       number = {1},
        pages = {29-47},
          doi = {10.1006/icar.2000.6585},
       adsurl = {https://ui.adsabs.harvard.edu/abs/2001Icar..152...29E}
}

@ARTICLE{Chambers_1999,
       author = {{Chambers}, J.~E.},
        title = "{A hybrid symplectic integrator that permits close encounters between massive bodies}",
      journal = {\mnras},
         year = 1999,
        month = apr,
       volume = {304},
       number = {4},
        pages = {793-799},
          doi = {10.1046/j.1365-8711.1999.02379.x},
       adsurl = {https://ui.adsabs.harvard.edu/abs/1999MNRAS.304..793C}
}

@BOOK{Murray+Dermott_1999,
       author = {{Murray}, Carl D. and {Dermott}, Stanley F.},
        title = "{Solar System Dynamics}",
         year = 1999,
          doi = {10.1017/CBO9781139174817},
       adsurl = {https://ui.adsabs.harvard.edu/abs/1999ssd..book.....M}
}

@ARTICLE{Smith+Lissauer_2009,
       author = {{Smith}, Andrew W. and {Lissauer}, Jack J.},
        title = "{Orbital stability of systems of closely-spaced planets}",
      journal = {\icarus},
         year = 2009,
        month = may,
       volume = {201},
       number = {1},
        pages = {381-394},
          doi = {10.1016/j.icarus.2008.12.027},
       adsurl = {https://ui.adsabs.harvard.edu/abs/2009Icar..201..381S}
}

@ARTICLE{Smith+Lissauer_2010,
       author = {{Smith}, Andrew W. and {Lissauer}, Jack J.},
        title = "{Orbital stability of systems of closely-spaced planets, II: configurations with coorbital planets}",
      journal = {Celestial Mechanics and Dynamical Astronomy},
         year = 2010,
        month = aug,
       volume = {107},
       number = {4},
        pages = {487-500},
          doi = {10.1007/s10569-010-9288-0},
       adsurl = {https://ui.adsabs.harvard.edu/abs/2010CeMDA.107..487S}
}

@ARTICLE{Quarles_2018,
       author = {{Quarles}, B. and {Lissauer}, Jack J.},
        title = "{Long-term Stability of Tightly Packed Multi-planet Systems in Prograde, Coplanar, Circumstellar Orbits within the {\ensuremath{\alpha}} Centauri AB System}",
      journal = {\aj},
         year = 2018,
        month = mar,
       volume = {155},
       number = {3},
          eid = {130},
        pages = {130},
          doi = {10.3847/1538-3881/aaa966},
archivePrefix = {arXiv},
       eprint = {1801.06131},
 primaryClass = {astro-ph.EP},
       adsurl = {https://ui.adsabs.harvard.edu/abs/2018AJ....155..130Q}
}

@ARTICLE{Lissauer+etal_2014,
       author = {{Lissauer}, Jack J. and {Marcy}, Geoffrey W. and {Bryson}, Stephen T. and {Rowe}, Jason F. and {Jontof-Hutter}, Daniel and {Agol}, Eric and {Borucki}, William J. and {Carter}, Joshua A. and {Ford}, Eric B. and {Gilliland}, Ronald L. and {Kolbl}, Rea and {Star}, Kimberly M. and {Steffen}, Jason H. and {Torres}, Guillermo},
        title = "{Validation of Kepler's Multiple Planet Candidates. II. Refined Statistical Framework and Descriptions of Systems of Special Interest}",
      journal = {\apj},
         year = 2014,
        month = mar,
       volume = {784},
       number = {1},
          eid = {44},
        pages = {44},
          doi = {10.1088/0004-637X/784/1/44},
archivePrefix = {arXiv},
       eprint = {1402.6352},
 primaryClass = {astro-ph.EP},
       adsurl = {https://ui.adsabs.harvard.edu/abs/2014ApJ...784...44L}
}

@ARTICLE{Quillen_2011,
       author = {{Quillen}, Alice C.},
        title = "{Three-body resonance overlap in closely spaced multiple-planet systems}",
      journal = {\mnras},
         year = 2011,
        month = dec,
       volume = {418},
       number = {2},
        pages = {1043-1054},
          doi = {10.1111/j.1365-2966.2011.19555.x},
archivePrefix = {arXiv},
       eprint = {1106.0156},
 primaryClass = {astro-ph.EP},
       adsurl = {https://ui.adsabs.harvard.edu/abs/2011MNRAS.418.1043Q}
}

@ARTICLE{Lissauer+etal_2011a,
       author = {{Lissauer}, Jack J. and {Fabrycky}, Daniel C. and {Ford}, Eric B. and {Borucki}, William J. and {Fressin}, Francois and {Marcy}, Geoffrey W. and {Orosz}, Jerome A. and {Rowe}, Jason F. and {Torres}, Guillermo and {Welsh}, William F. and {Batalha}, Natalie M. and {Bryson}, Stephen T. and {Buchhave}, Lars A. and {Caldwell}, Douglas A. and {Carter}, Joshua A. and {Charbonneau}, David and {Christiansen}, Jessie L. and {Cochran}, William D. and {Desert}, Jean-Michel and {Dunham}, Edward W. and {Fanelli}, Michael N. and {Fortney}, Jonathan J. and {Gautier}, Thomas N., III and {Geary}, John C. and {Gilliland}, Ronald L. and {Haas}, Michael R. and {Hall}, Jennifer R. and {Holman}, Matthew J. and {Koch}, David G. and {Latham}, David W. and {Lopez}, Eric and {McCauliff}, Sean and {Miller}, Neil and {Morehead}, Robert C. and {Quintana}, Elisa V. and {Ragozzine}, Darin and {Sasselov}, Dimitar and {Short}, Donald R. and {Steffen}, Jason H.},
        title = "{A closely packed system of low-mass, low-density planets transiting Kepler-11}",
      journal = {\nat},
         year = 2011,
        month = feb,
       volume = {470},
       number = {7332},
        pages = {53-58},
          doi = {10.1038/nature09760},
archivePrefix = {arXiv},
       eprint = {1102.0291},
 primaryClass = {astro-ph.EP},
       adsurl = {https://ui.adsabs.harvard.edu/abs/2011Natur.470...53L}
}

@ARTICLE{Lissauer+etal_2011b,
       author = {{Lissauer}, Jack J. and {Ragozzine}, Darin and {Fabrycky}, Daniel C. and {Steffen}, Jason H. and {Ford}, Eric B. and {Jenkins}, Jon M. and {Shporer}, Avi and {Holman}, Matthew J. and {Rowe}, Jason F. and {Quintana}, Elisa V. and {Batalha}, Natalie M. and {Borucki}, William J. and {Bryson}, Stephen T. and {Caldwell}, Douglas A. and {Carter}, Joshua A. and {Ciardi}, David and {Dunham}, Edward W. and {Fortney}, Jonathan J. and {Gautier}, Thomas N., III and {Howell}, Steve B. and {Koch}, David G. and {Latham}, David W. and {Marcy}, Geoffrey W. and {Morehead}, Robert C. and {Sasselov}, Dimitar},
        title = "{Architecture and Dynamics of Kepler's Candidate Multiple Transiting Planet Systems}",
      journal = {\apjs},
         year = 2011,
        month = nov,
       volume = {197},
       number = {1},
          eid = {8},
        pages = {8},
          doi = {10.1088/0067-0049/197/1/8},
archivePrefix = {arXiv},
       eprint = {1102.0543},
 primaryClass = {astro-ph.EP},
       adsurl = {https://ui.adsabs.harvard.edu/abs/2011ApJS..197....8L}
}

@ARTICLE{Deck+etal_2013,
       author = {{Deck}, Katherine M. and {Payne}, Matthew and {Holman}, Matthew J.},
        title = "{First-order Resonance Overlap and the Stability of Close Two-planet Systems}",
      journal = {\apj},
         year = 2013,
        month = sep,
       volume = {774},
       number = {2},
          eid = {129},
        pages = {129},
          doi = {10.1088/0004-637X/774/2/129},
archivePrefix = {arXiv},
       eprint = {1307.8119},
 primaryClass = {astro-ph.EP},
       adsurl = {https://ui.adsabs.harvard.edu/abs/2013ApJ...774..129D}
}

@ARTICLE{Steffen+etal_2013,
       author = {{Steffen}, Jason H. and {Fabrycky}, Daniel C. and {Agol}, Eric and {Ford}, Eric B. and {Morehead}, Robert C. and {Cochran}, William D. and {Lissauer}, Jack J. and {Adams}, Elisabeth R. and {Borucki}, William J. and {Bryson}, Steve and {Caldwell}, Douglas A. and {Dupree}, Andrea and {Jenkins}, Jon M. and {Robertson}, Paul and {Rowe}, Jason F. and {Seader}, Shawn and {Thompson}, Susan and {Twicken}, Joseph D.},
        title = "{Transit timing observations from Kepler - VII. Confirmation of 27 planets in 13 multiplanet systems via transit timing variations and orbital stability}",
      journal = {\mnras},
         year = 2013,
        month = jan,
       volume = {428},
       number = {2},
        pages = {1077-1087},
          doi = {10.1093/mnras/sts090},
archivePrefix = {arXiv},
       eprint = {1208.3499},
 primaryClass = {astro-ph.EP},
       adsurl = {https://ui.adsabs.harvard.edu/abs/2013MNRAS.428.1077S}
}

@ARTICLE{Fabrycky+etal_2014,
       author = {{Fabrycky}, Daniel C. and {Lissauer}, Jack J. and {Ragozzine}, Darin and {Rowe}, Jason F. and {Steffen}, Jason H. and {Agol}, Eric and {Barclay}, Thomas and {Batalha}, Natalie and {Borucki}, William and {Ciardi}, David R. and {Ford}, Eric B. and {Gautier}, Thomas N. and {Geary}, John C. and {Holman}, Matthew J. and {Jenkins}, Jon M. and {Li}, Jie and {Morehead}, Robert C. and {Morris}, Robert L. and {Shporer}, Avi and {Smith}, Jeffrey C. and {Still}, Martin and {Van Cleve}, Jeffrey},
        title = "{Architecture of Kepler's Multi-transiting Systems. II. New Investigations with Twice as Many Candidates}",
      journal = {\apj},
         year = 2014,
        month = aug,
       volume = {790},
       number = {2},
          eid = {146},
        pages = {146},
          doi = {10.1088/0004-637X/790/2/146},
archivePrefix = {arXiv},
       eprint = {1202.6328},
 primaryClass = {astro-ph.EP},
       adsurl = {https://ui.adsabs.harvard.edu/abs/2014ApJ...790..146F}
}

@ARTICLE{hands+etal_2014,
       author = {{Hands}, T.~O. and {Alexander}, R.~D. and {Dehnen}, W.},
        title = "{Understanding the assembly of Kepler's compact planetary systems}",
      journal = {\mnras},
         year = 2014,
        month = nov,
       volume = {445},
       number = {1},
        pages = {749-760},
          doi = {10.1093/mnras/stu1751},
archivePrefix = {arXiv},
       eprint = {1409.0532},
 primaryClass = {astro-ph.EP},
       adsurl = {https://ui.adsabs.harvard.edu/abs/2014MNRAS.445..749H}
}

@ARTICLE{Rowe+etal_2014,
       author = {{Rowe}, Jason F. and {Bryson}, Stephen T. and {Marcy}, Geoffrey W. and {Lissauer}, Jack J. and {Jontof-Hutter}, Daniel and {Mullally}, Fergal and {Gilliland}, Ronald L. and {Issacson}, Howard and {Ford}, Eric and {Howell}, Steve B. and {Borucki}, William J. and {Haas}, Michael and {Huber}, Daniel and {Steffen}, Jason H. and {Thompson}, Susan E. and {Quintana}, Elisa and {Barclay}, Thomas and {Still}, Martin and {Fortney}, Jonathan and {Gautier}, III, T.~N. and {Hunter}, Roger and {Caldwell}, Douglas A. and {Ciardi}, David R. and {Devore}, Edna and {Cochran}, William and {Jenkins}, Jon and {Agol}, Eric and {Carter}, Joshua A. and {Geary}, John},
        title = "{Validation of Kepler's Multiple Planet Candidates. III. Light Curve Analysis and Announcement of Hundreds of New Multi-planet Systems}",
      journal = {\apj},
         year = 2014,
        month = mar,
       volume = {784},
       number = {1},
          eid = {45},
        pages = {45},
          doi = {10.1088/0004-637X/784/1/45},
archivePrefix = {arXiv},
       eprint = {1402.6534},
 primaryClass = {astro-ph.EP},
       adsurl = {https://ui.adsabs.harvard.edu/abs/2014ApJ...784...45R}
}

@ARTICLE{Papaloizou+etal_2015,
       author = {{Papaloizou}, John C.~B.},
        title = "{Three body resonances in close orbiting planetary systems: tidal dissipation and orbital evolution}",
      journal = {International Journal of Astrobiology},
         year = 2015,
        month = apr,
       volume = {14},
       number = {2},
        pages = {291-304},
          doi = {10.1017/S1473550414000147},
archivePrefix = {arXiv},
       eprint = {1405.0381},
 primaryClass = {astro-ph.EP},
       adsurl = {https://ui.adsabs.harvard.edu/abs/2015IJAsB..14..291P}
}

@ARTICLE{Morrison+Kratter_2016,
       author = {{Morrison}, Sarah J. and {Kratter}, Kaitlin M.},
        title = "{Orbital Stability of Multi-planet Systems: Behavior at High Masses}",
      journal = {\apj},
         year = 2016,
        month = jun,
       volume = {823},
       number = {2},
          eid = {118},
        pages = {118},
          doi = {10.3847/0004-637X/823/2/118},
archivePrefix = {arXiv},
       eprint = {1604.01037},
 primaryClass = {astro-ph.EP},
       adsurl = {https://ui.adsabs.harvard.edu/abs/2016ApJ...823..118M}
}

@ARTICLE{Gozd+etal_2016,
       author = {{Go{\'z}dziewski}, K. and {Migaszewski}, C. and {Panichi}, F. and {Szuszkiewicz}, E.},
        title = "{The Laplace resonance in the Kepler-60 planetary system}",
      journal = {\mnras},
         year = 2016,
        month = jan,
       volume = {455},
       number = {1},
        pages = {L104-L108},
          doi = {10.1093/mnrasl/slv156},
archivePrefix = {arXiv},
       eprint = {1510.02776},
 primaryClass = {astro-ph.EP},
       adsurl = {https://ui.adsabs.harvard.edu/abs/2016MNRAS.455L.104G}
}

@ARTICLE{MacDonald+etal_2016,
       author = {{MacDonald}, Mariah G. and {Ragozzine}, Darin and {Fabrycky}, Daniel C. and {Ford}, Eric B. and {Holman}, Matthew J. and {Isaacson}, Howard T. and {Lissauer}, Jack J. and {Lopez}, Eric D. and {Mazeh}, Tsevi and {Rogers}, Leslie and {Rowe}, Jason F. and {Steffen}, Jason H. and {Torres}, Guillermo},
        title = "{A Dynamical Analysis of the Kepler-80 System of Five Transiting Planets}",
      journal = {\aj},
         year = 2016,
        month = oct,
       volume = {152},
       number = {4},
          eid = {105},
        pages = {105},
          doi = {10.3847/0004-6256/152/4/105},
archivePrefix = {arXiv},
       eprint = {1607.07540},
 primaryClass = {astro-ph.EP},
       adsurl = {https://ui.adsabs.harvard.edu/abs/2016AJ....152..105M}
}

@ARTICLE{Jontof+etal_2016,
       author = {{Jontof-Hutter}, Daniel and {Ford}, Eric B. and {Rowe}, Jason F. and {Lissauer}, Jack J. and {Fabrycky}, Daniel C. and {Van Laerhoven}, Christa and {Agol}, Eric and {Deck}, Katherine M. and {Holczer}, Tomer and {Mazeh}, Tsevi},
        title = "{Secure Mass Measurements from Transit Timing: 10 Kepler Exoplanets between 3 and 8 M$_{{\ensuremath{\oplus}}}$ with Diverse Densities and Incident Fluxes}",
      journal = {\apj},
         year = 2016,
        month = mar,
       volume = {820},
       number = {1},
          eid = {39},
        pages = {39},
          doi = {10.3847/0004-637X/820/1/39},
archivePrefix = {arXiv},
       eprint = {1512.02003},
 primaryClass = {astro-ph.EP},
       adsurl = {https://ui.adsabs.harvard.edu/abs/2016ApJ...820...39J}
}

@ARTICLE{Mills+etal_2016,
       author = {{Mills}, Sean M. and {Fabrycky}, Daniel C. and {Migaszewski}, Cezary and {Ford}, Eric B. and {Petigura}, Erik and {Isaacson}, Howard},
        title = "{A resonant chain of four transiting, sub-Neptune planets}",
      journal = {\nat},
         year = 2016,
        month = may,
       volume = {533},
       number = {7604},
        pages = {509-512},
          doi = {10.1038/nature17445},
archivePrefix = {arXiv},
       eprint = {1612.07376},
 primaryClass = {astro-ph.EP},
       adsurl = {https://ui.adsabs.harvard.edu/abs/2016Natur.533..509M}
}

@ARTICLE{tamayo+etal_2016,
       author = {{Tamayo}, Daniel and {Silburt}, Ari and {Valencia}, Diana and {Menou}, Kristen and {Ali-Dib}, Mohamad and {Petrovich}, Cristobal and {Huang}, Chelsea X. and {Rein}, Hanno and {van Laerhoven}, Christa and {Paradise}, Adiv and {Obertas}, Alysa and {Murray}, Norman},
        title = "{A Machine Learns to Predict the Stability of Tightly Packed Planetary Systems}",
      journal = {\apjl},
         year = 2016,
        month = dec,
       volume = {832},
       number = {2},
          eid = {L22},
        pages = {L22},
          doi = {10.3847/2041-8205/832/2/L22},
archivePrefix = {arXiv},
       eprint = {1610.05359},
 primaryClass = {astro-ph.EP},
       adsurl = {https://ui.adsabs.harvard.edu/abs/2016ApJ...832L..22T}
}

@ARTICLE{Obertas+etal_2017,
       author = {{Obertas}, Alysa and {Van Laerhoven}, Christa and {Tamayo}, Daniel},
        title = "{The stability of tightly-packed, evenly-spaced systems of Earth-mass planets orbiting a Sun-like star}",
      journal = {\icarus},
         year = 2017,
        month = sep,
       volume = {293},
        pages = {52-58},
          doi = {10.1016/j.icarus.2017.04.010},
archivePrefix = {arXiv},
       eprint = {1703.08426},
 primaryClass = {astro-ph.EP},
       adsurl = {https://ui.adsabs.harvard.edu/abs/2017Icar..293...52O}
}

@ARTICLE{Batygin+Adams_2017,
       author = {{Batygin}, Konstantin and {Adams}, Fred C.},
        title = "{An Analytic Criterion for Turbulent Disruption of Planetary Resonances}",
      journal = {\aj},
         year = 2017,
        month = mar,
       volume = {153},
       number = {3},
          eid = {120},
        pages = {120},
          doi = {10.3847/1538-3881/153/3/120},
archivePrefix = {arXiv},
       eprint = {1701.07849},
 primaryClass = {astro-ph.EP},
       adsurl = {https://ui.adsabs.harvard.edu/abs/2017AJ....153..120B}
}

@ARTICLE{Guillon+2017,
       author = {{Gillon}, Micha{\"e}l and {Triaud}, Amaury H.~M.~J. and {Demory}, Brice-Olivier and {Jehin}, Emmanu{\"e}l and {Agol}, Eric and {Deck}, Katherine M. and {Lederer}, Susan M. and {de Wit}, Julien and {Burdanov}, Artem and {Ingalls}, James G. and {Bolmont}, Emeline and {Leconte}, Jeremy and {Raymond}, Sean N. and {Selsis}, Franck and {Turbet}, Martin and {Barkaoui}, Khalid and {Burgasser}, Adam and {Burleigh}, Matthew R. and {Carey}, Sean J. and {Chaushev}, Aleksander and {Copperwheat}, Chris M. and {Delrez}, Laetitia and {Fernandes}, Catarina S. and {Holdsworth}, Daniel L. and {Kotze}, Enrico J. and {Van Grootel}, Val{\'e}rie and {Almleaky}, Yaseen and {Benkhaldoun}, Zouhair and {Magain}, Pierre and {Queloz}, Didier},
        title = "{Seven temperate terrestrial planets around the nearby ultracool dwarf star TRAPPIST-1}",
      journal = {\nat},
         year = 2017,
        month = feb,
       volume = {542},
       number = {7642},
        pages = {456-460},
          doi = {10.1038/nature21360},
archivePrefix = {arXiv},
       eprint = {1703.01424},
 primaryClass = {astro-ph.EP},
       adsurl = {https://ui.adsabs.harvard.edu/abs/2017Natur.542..456G}
}

@ARTICLE{Luger+etal_2017,
       author = {{Luger}, Rodrigo and {Sestovic}, Marko and {Kruse}, Ethan and {Grimm}, Simon L. and {Demory}, Brice-Olivier and {Agol}, Eric and {Bolmont}, Emeline and {Fabrycky}, Daniel and {Fernandes}, Catarina S. and {Van Grootel}, Val{\'e}rie and {Burgasser}, Adam and {Gillon}, Micha{\"e}l and {Ingalls}, James G. and {Jehin}, Emmanu{\"e}l and {Raymond}, Sean N. and {Selsis}, Franck and {Triaud}, Amaury H.~M.~J. and {Barclay}, Thomas and {Barentsen}, Geert and {Howell}, Steve B. and {Delrez}, Laetitia and {de Wit}, Julien and {Foreman-Mackey}, Daniel and {Holdsworth}, Daniel L. and {Leconte}, J{\'e}r{\'e}my and {Lederer}, Susan and {Turbet}, Martin and {Almleaky}, Yaseen and {Benkhaldoun}, Zouhair and {Magain}, Pierre and {Morris}, Brett M. and {Heng}, Kevin and {Queloz}, Didier},
        title = "{A seven-planet resonant chain in TRAPPIST-1}",
      journal = {Nature Astronomy},
         year = 2017,
        month = jun,
       volume = {1},
          eid = {0129},
        pages = {0129},
          doi = {10.1038/s41550-017-0129},
archivePrefix = {arXiv},
       eprint = {1703.04166},
 primaryClass = {astro-ph.EP},
       adsurl = {https://ui.adsabs.harvard.edu/abs/2017NatAs...1E.129L}
}

@ARTICLE{Millholland+etal_2017,
       author = {{Millholland}, Sarah and {Wang}, Songhu and {Laughlin}, Gregory},
        title = "{Kepler Multi-planet Systems Exhibit Unexpected Intra-system Uniformity in Mass and Radius}",
      journal = {\apjl},
         year = 2017,
        month = nov,
       volume = {849},
       number = {2},
          eid = {L33},
        pages = {L33},
          doi = {10.3847/2041-8213/aa9714},
archivePrefix = {arXiv},
       eprint = {1710.11152},
 primaryClass = {astro-ph.EP},
       adsurl = {https://ui.adsabs.harvard.edu/abs/2017ApJ...849L..33M}
}

@ARTICLE{Izidoro+etal_2017,
       author = {{Izidoro}, Andre and {Ogihara}, Masahiro and {Raymond}, Sean N. and {Morbidelli}, Alessandro and {Pierens}, Arnaud and {Bitsch}, Bertram and {Cossou}, Christophe and {Hersant}, Franck},
        title = "{Breaking the chains: hot super-Earth systems from migration and disruption of compact resonant chains}",
      journal = {\mnras},
         year = 2017,
        month = sep,
       volume = {470},
       number = {2},
        pages = {1750-1770},
          doi = {10.1093/mnras/stx1232},
archivePrefix = {arXiv},
       eprint = {1703.03634},
 primaryClass = {astro-ph.EP},
       adsurl = {https://ui.adsabs.harvard.edu/abs/2017MNRAS.470.1750I}
}

@ARTICLE{Hadden+Lithwick_2018,
       author = {{Hadden}, Sam and {Lithwick}, Yoram},
        title = "{A Criterion for the Onset of Chaos in Systems of Two Eccentric Planets}",
      journal = {\aj},
         year = 2018,
        month = sep,
       volume = {156},
       number = {3},
          eid = {95},
        pages = {95},
          doi = {10.3847/1538-3881/aad32c},
archivePrefix = {arXiv},
       eprint = {1803.08510},
 primaryClass = {astro-ph.EP},
       adsurl = {https://ui.adsabs.harvard.edu/abs/2018AJ....156...95H}
}

@ARTICLE{Weiss+etal_2018,
       author = {{Weiss}, Lauren M. and {Marcy}, Geoffrey W. and {Petigura}, Erik A. and {Fulton}, Benjamin J. and {Howard}, Andrew W. and {Winn}, Joshua N. and {Isaacson}, Howard T. and {Morton}, Timothy D. and {Hirsch}, Lea A. and {Sinukoff}, Evan J. and {Cumming}, Andrew and {Hebb}, Leslie and {Cargile}, Phillip A.},
        title = "{The California-Kepler Survey. V. Peas in a Pod: Planets in a Kepler Multi-planet System Are Similar in Size and Regularly Spaced}",
      journal = {\aj},
         year = 2018,
        month = jan,
       volume = {155},
       number = {1},
          eid = {48},
        pages = {48},
          doi = {10.3847/1538-3881/aa9ff6},
archivePrefix = {arXiv},
       eprint = {1706.06204},
 primaryClass = {astro-ph.EP},
       adsurl = {https://ui.adsabs.harvard.edu/abs/2018AJ....155...48W}
}

@ARTICLE{Petit+etal_2018,
       author = {{Petit}, Antoine C. and {Laskar}, Jacques and {Bou{\'e}}, Gwena{\"e}l},
        title = "{Hill stability in the AMD framework}",
      journal = {\aap},
         year = 2018,
        month = sep,
       volume = {617},
          eid = {A93},
        pages = {A93},
          doi = {10.1051/0004-6361/201833088},
archivePrefix = {arXiv},
       eprint = {1806.08869},
 primaryClass = {astro-ph.EP},
       adsurl = {https://ui.adsabs.harvard.edu/abs/2018A&A...617A..93P}
}

@ARTICLE{Shallue+_2018,
       author = {{Shallue}, Christopher J. and {Vanderburg}, Andrew},
        title = "{Identifying Exoplanets with Deep Learning: A Five-planet Resonant Chain around Kepler-80 and an Eighth Planet around Kepler-90}",
      journal = {\aj},
         year = 2018,
        month = feb,
       volume = {155},
       number = {2},
          eid = {94},
        pages = {94},
          doi = {10.3847/1538-3881/aa9e09},
archivePrefix = {arXiv},
       eprint = {1712.05044},
 primaryClass = {astro-ph.EP},
       adsurl = {https://ui.adsabs.harvard.edu/abs/2018AJ....155...94S}
}

@ARTICLE{Petit+etal_2020,
       author = {{Petit}, Antoine C. and {Pichierri}, Gabriele and {Davies}, Melvyn B. and {Johansen}, Anders},
        title = "{The path to instability in compact multi-planetary systems}",
      journal = {\aap},
         year = 2020,
        month = sep,
       volume = {641},
          eid = {A176},
        pages = {A176},
          doi = {10.1051/0004-6361/202038764},
archivePrefix = {arXiv},
       eprint = {2006.14903},
 primaryClass = {astro-ph.EP},
       adsurl = {https://ui.adsabs.harvard.edu/abs/2020A&A...641A.176P}
}

@ARTICLE{Pichierri+Morbidelli_2020,
       author = {{Pichierri}, Gabriele and {Morbidelli}, Alessandro},
        title = "{The onset of instability in resonant chains}",
      journal = {\mnras},
         year = 2020,
        month = jun,
       volume = {494},
       number = {4},
        pages = {4950-4968},
          doi = {10.1093/mnras/staa1102},
archivePrefix = {arXiv},
       eprint = {2004.07789},
 primaryClass = {astro-ph.EP},
       adsurl = {https://ui.adsabs.harvard.edu/abs/2020MNRAS.494.4950P}
}

@ARTICLE{Lissauer&Gavino_2021,
       author = {{Lissauer}, Jack J. and {Gavino}, Sacha},
        title = "{Orbital stability of compact three-planet systems, I: Dependence of system lifetimes on initial orbital separations and longitudes}",
      journal = {\icarus},
         year = 2021,
        month = aug,
       volume = {364},
          eid = {114470},
        pages = {114470},
          doi = {10.1016/j.icarus.2021.114470},
archivePrefix = {arXiv},
       eprint = {2104.13657},
 primaryClass = {astro-ph.EP},
       adsurl = {https://ui.adsabs.harvard.edu/abs/2021Icar..36414470L}
}

@ARTICLE{Bartram+etal_2021,
       author = {{Bartram}, Peter and {Wittig}, Alexander and {Lissauer}, Jack J. and {Gavino}, Sacha and {Urrutxua}, Hodei},
        title = "{Orbital stability of compact three-planet systems - II: post-instability impact behaviour}",
      journal = {\mnras},
         year = 2021,
        month = oct,
       volume = {506},
       number = {4},
        pages = {6181-6194},
          doi = {10.1093/mnras/stab1465},
archivePrefix = {arXiv},
       eprint = {2104.13658},
 primaryClass = {astro-ph.EP},
       adsurl = {https://ui.adsabs.harvard.edu/abs/2021MNRAS.506.6181B}
}

@ARTICLE{Petit+etal_2021,
       author = {{Petit}, Antoine C.},
        title = "{An integrable model for first-order three-planet mean motion resonances}",
      journal = {Celestial Mechanics and Dynamical Astronomy},
         year = 2021,
        month = aug,
       volume = {133},
       number = {8},
          eid = {39},
        pages = {39},
          doi = {10.1007/s10569-021-10035-7},
archivePrefix = {arXiv},
       eprint = {2107.06299},
 primaryClass = {astro-ph.EP},
       adsurl = {https://ui.adsabs.harvard.edu/abs/2021CeMDA.133...39P}
}

@ARTICLE{MacDonald+etal_2021,
       author = {{MacDonald}, Mariah G. and {Shakespeare}, Cody J. and {Ragozzine}, Darin},
        title = "{A Five-Planet Resonant Chain: Reevaluation of the Kepler-80 System}",
      journal = {\aj},
         year = 2021,
        month = sep,
       volume = {162},
       number = {3},
          eid = {114},
        pages = {114},
          doi = {10.3847/1538-3881/ac12d5},
archivePrefix = {arXiv},
       eprint = {2107.05597},
 primaryClass = {astro-ph.EP},
       adsurl = {https://ui.adsabs.harvard.edu/abs/2021AJ....162..114M}
}

@ARTICLE{Izidoro+etal_2021,
       author = {{Izidoro}, Andr{\'e} and {Bitsch}, Bertram and {Raymond}, Sean N. and {Johansen}, Anders and {Morbidelli}, Alessandro and {Lambrechts}, Michiel and {Jacobson}, Seth A.},
        title = "{Formation of planetary systems by pebble accretion and migration. Hot super-Earth systems from breaking compact resonant chains}",
      journal = {\aap},
         year = 2021,
        month = jun,
       volume = {650},
          eid = {A152},
        pages = {A152},
          doi = {10.1051/0004-6361/201935336},
archivePrefix = {arXiv},
       eprint = {1902.08772},
 primaryClass = {astro-ph.EP},
       adsurl = {https://ui.adsabs.harvard.edu/abs/2021A&A...650A.152I}
}

@ARTICLE{Agol+etal_2021,
       author = {{Agol}, Eric and {Dorn}, Caroline and {Grimm}, Simon L. and {Turbet}, Martin and {Ducrot}, Elsa and {Delrez}, Laetitia and {Gillon}, Micha{\"e}l and {Demory}, Brice-Olivier and {Burdanov}, Artem and {Barkaoui}, Khalid and {Benkhaldoun}, Zouhair and {Bolmont}, Emeline and {Burgasser}, Adam and {Carey}, Sean and {de Wit}, Julien and {Fabrycky}, Daniel and {Foreman-Mackey}, Daniel and {Haldemann}, Jonas and {Hernandez}, David M. and {Ingalls}, James and {Jehin}, Emmanuel and {Langford}, Zachary and {Leconte}, J{\'e}r{\'e}my and {Lederer}, Susan M. and {Luger}, Rodrigo and {Malhotra}, Renu and {Meadows}, Victoria S. and {Morris}, Brett M. and {Pozuelos}, Francisco J. and {Queloz}, Didier and {Raymond}, Sean N. and {Selsis}, Franck and {Sestovic}, Marko and {Triaud}, Amaury H.~M.~J. and {Van Grootel}, Valerie},
        title = "{Refining the Transit-timing and Photometric Analysis of TRAPPIST-1: Masses, Radii, Densities, Dynamics, and Ephemerides}",
      journal = {PSJ},
         year = 2021,
        month = feb,
       volume = {2},
       number = {1},
          eid = {1},
        pages = {1},
          doi = {10.3847/PSJ/abd022},
archivePrefix = {arXiv},
       eprint = {2010.01074},
 primaryClass = {astro-ph.EP},
       adsurl = {https://ui.adsabs.harvard.edu/abs/2021PSJ.....2....1A}
}

@ARTICLE{Goldberg+Batygin_2022,
       author = {{Goldberg}, Max and {Batygin}, Konstantin},
        title = "{Architectures of Compact Super-Earth Systems Shaped by Instabilities}",
      journal = {\aj},
         year = 2022,
        month = may,
       volume = {163},
       number = {5},
          eid = {201},
        pages = {201},
          doi = {10.3847/1538-3881/ac5961},
archivePrefix = {arXiv},
       eprint = {2203.00801},
 primaryClass = {astro-ph.EP},
       adsurl = {https://ui.adsabs.harvard.edu/abs/2022AJ....163..201G}
}

@ARTICLE{Weiss+etal_2023,
       author = {{Weiss}, L.~M. and {Millholland}, S.~C. and {Petigura}, E.~A. and {Adams}, F.~C. and {Batygin}, K. and {Block}, A.~M. and {Mordasini}, C.},
        title = "{Architectures of Compact Multi-Planet Systems: Diversity and Uniformity}",
    booktitle = {Protostars and Planets VII},
         year = 2023,
       editor = {{Inutsuka}, S. and {Aikawa}, Y. and {Muto}, T. and {Tomida}, K. and {Tamura}, M.},
       series = {Astronomical Society of the Pacific Conference Series},
       volume = {534},
        month = jul,
        pages = {863},
          doi = {10.48550/arXiv.2203.10076},
archivePrefix = {arXiv},
       eprint = {2203.10076},
 primaryClass = {astro-ph.EP},
       adsurl = {https://ui.adsabs.harvard.edu/abs/2023ASPC..534..863W}
}

@ARTICLE{Tejada_Arevalo+etal_2022,
       author = {{Tejada Arevalo}, Roberto and {Tamayo}, Daniel and {Cranmer}, Miles},
        title = "{Stability Constrained Characterization of the 23 Myr Old V1298 Tau System: Do Young Planets Form in Mean Motion Resonance Chains?}",
      journal = {\apjl},
         year = 2022,
        month = jun,
       volume = {932},
       number = {1},
          eid = {L12},
        pages = {L12},
          doi = {10.3847/2041-8213/ac70e0},
archivePrefix = {arXiv},
       eprint = {2203.02805},
 primaryClass = {astro-ph.EP},
       adsurl = {https://ui.adsabs.harvard.edu/abs/2022ApJ...932L..12T}
}

@ARTICLE{Cerioni+etal_2022,
       author = {{Cerioni}, M. and {Beaug{\'e}}, C. and {Gallardo}, T.},
        title = "{Is the orbital distribution of multiplanet systems influenced by pure three-planet resonances?}",
      journal = {\mnras},
         year = 2022,
        month = jun,
       volume = {513},
       number = {1},
        pages = {541-550},
          doi = {10.1093/mnras/stac876},
archivePrefix = {arXiv},
       eprint = {2204.12364},
 primaryClass = {astro-ph.EP},
       adsurl = {https://ui.adsabs.harvard.edu/abs/2022MNRAS.513..541C}
}

@ARTICLE{Rice+Steffen_2023,
       author = {{Rice}, David R. and {Steffen}, Jason H.},
        title = "{Stable lifetime of compact, evenly spaced planetary systems with non-equal masses}",
      journal = {\mnras},
         year = 2023,
        month = apr,
       volume = {520},
       number = {3},
        pages = {4057-4067},
          doi = {10.1093/mnras/stad393},
archivePrefix = {arXiv},
       eprint = {2206.11374},
 primaryClass = {astro-ph.EP},
       adsurl = {https://ui.adsabs.harvard.edu/abs/2023MNRAS.520.4057R}
}

@ARTICLE{Tamayo+Hadden_2025,
       author = {{Tamayo}, Daniel and {Hadden}, Samuel},
        title = "{A Unified, Physical Framework for Mean Motion Resonances}",
      journal = {\apj},
         year = 2025,
        month = jun,
       volume = {986},
       number = {1},
          eid = {11},
        pages = {11},
          doi = {10.3847/1538-4357/adc1c4},
archivePrefix = {arXiv},
       eprint = {2410.21748},
 primaryClass = {astro-ph.EP},
       adsurl = {https://ui.adsabs.harvard.edu/abs/2025ApJ...986...11T}
}

@ARTICLE{Rebound,
       author = {{Rein}, H. and {Liu}, S. -F.},
        title = "{REBOUND: an open-source multi-purpose N-body code for collisional dynamics}",
      journal = {\aap},
         year = 2012,
        month = jan,
       volume = {537},
          eid = {A128},
        pages = {A128},
          doi = {10.1051/0004-6361/201118085},
archivePrefix = {arXiv},
       eprint = {1110.4876},
 primaryClass = {astro-ph.EP},
       adsurl = {https://ui.adsabs.harvard.edu/abs/2012A&A...537A.128R}
}

@ARTICLE{reboundwhfast,
       author = {{Rein}, Hanno and {Tamayo}, Daniel},
        title = "{WHFAST: a fast and unbiased implementation of a symplectic Wisdom-Holman integrator for long-term gravitational simulations}",
      journal = {\mnras},
         year = 2015,
        month = sep,
       volume = {452},
       number = {1},
        pages = {376-388},
          doi = {10.1093/mnras/stv1257},
archivePrefix = {arXiv},
       eprint = {1506.01084},
 primaryClass = {astro-ph.EP},
       adsurl = {https://ui.adsabs.harvard.edu/abs/2015MNRAS.452..376R}
}

@ARTICLE{wh,
       author = {{Wisdom}, Jack and {Holman}, Matthew},
        title = "{Symplectic maps for the N-body problem.}",
      journal = {\aj},
         year = 1991,
        month = oct,
       volume = {102},
        pages = {1528-1538},
          doi = {10.1086/115978},
       adsurl = {https://ui.adsabs.harvard.edu/abs/1991AJ....102.1528W}
}

@ARTICLE{Lammers+etal_2024,
       author = {{Lammers}, Caleb and {Hadden}, Sam and {Murray}, Norman},
        title = "{The Instability Mechanism of Compact Multiplanet Systems}",
      journal = {\apj},
         year = 2024,
        month = sep,
       volume = {972},
       number = {1},
          eid = {53},
        pages = {53},
          doi = {10.3847/1538-4357/ad5be6},
archivePrefix = {arXiv},
       eprint = {2403.17928},
 primaryClass = {astro-ph.EP},
       adsurl = {https://ui.adsabs.harvard.edu/abs/2024ApJ...972...53L}
}

@ARTICLE{Gratia+Lissauer_2021,
       author = {{Gratia}, Pierre and {Lissauer}, Jack J.},
        title = "{Eccentricities and the stability of closely-spaced five-planet systems}",
      journal = {\icarus},
         year = 2021,
        month = apr,
       volume = {358},
          eid = {114038},
        pages = {114038},
          doi = {10.1016/j.icarus.2020.114038},
       adsurl = {https://ui.adsabs.harvard.edu/abs/2021Icar..35814038G}
}

@ARTICLE{Dai_2024,
       author = {{Dai}, Fei and {Goldberg}, Max and {Batygin}, Konstantin and {van Saders}, Jennifer and {Chiang}, Eugene and {Choksi}, Nick and {Li}, Rixin and {Petigura}, Erik A. and {Gilbert}, Gregory J. and {Millholland}, Sarah C. and {Dai}, Yuan-Zhe and {Bouma}, Luke and {Weiss}, Lauren M. and {Winn}, Joshua N.},
        title = "{The Prevalence of Resonance Among Young, Close-in Planets}",
      journal = {\aj},
         year = 2024,
        month = dec,
       volume = {168},
       number = {6},
          eid = {239},
        pages = {239},
          doi = {10.3847/1538-3881/ad83a6},
archivePrefix = {arXiv},
       eprint = {2406.06885},
 primaryClass = {astro-ph.EP},
       adsurl = {https://ui.adsabs.harvard.edu/abs/2024AJ....168..239D}
}

@ARTICLE{Lissauer+etal_2024,
       author = {{Lissauer}, Jack J. and {Rowe}, Jason F. and {Jontof-Hutter}, Daniel and {Fabrycky}, Daniel C. and {Ford}, Eric B. and {Ragozzine}, Darin and {Steffen}, Jason H. and {Nizam}, Kadri M.},
        title = "{Updated Catalog of Kepler Planet Candidates: Focus on Accuracy and Orbital Periods}",
      journal = {PSJ},
         year = 2024,
        month = jun,
       volume = {5},
       number = {6},
          eid = {152},
        pages = {152},
          doi = {10.3847/PSJ/ad0e6e},
archivePrefix = {arXiv},
       eprint = {2311.00238},
 primaryClass = {astro-ph.EP},
       adsurl = {https://ui.adsabs.harvard.edu/abs/2024PSJ.....5..152L}
}

@ARTICLE{Thadhani+etal_2025,
       author = {{Thadhani}, Elio and {Ba}, Yanming and {Rein}, Hanno and {Tamayo}, Daniel},
        title = "{SPOCK 2.0: Updates to the FeatureClassifier in the Stability of Planetary Orbital Configurations Klassifier}",
      journal = {Research Notes of the American Astronomical Society},
         year = 2025,
        month = feb,
       volume = {9},
       number = {2},
          eid = {27},
        pages = {27},
          doi = {10.3847/2515-5172/adb150},
archivePrefix = {arXiv},
       eprint = {2501.15017},
 primaryClass = {astro-ph.EP},
       adsurl = {https://ui.adsabs.harvard.edu/abs/2025RNAAS...9...27T}
}

@ARTICLE{Outland+etal_2026,
       author = {{Outland}, Bennet and {Noble}, Gretchen and {Smith}, Andrew W. and {Lissauer}, Jack J.},
        title = "{Orbital Stability of Closely-Spaced Four-planet Systems}",
      journal = {arXiv e-prints},
         year = 2026,
        month = jan,
          eid = {arXiv:2601.11692},
        pages = {arXiv:2601.11692},
archivePrefix = {arXiv},
       eprint = {2601.11692},
 primaryClass = {astro-ph.EP},
       adsurl = {https://ui.adsabs.harvard.edu/abs/2026arXiv260111692O}
}

\bibliographystyle{aa}

\begin{appendix}

\section{Stability of a system on the $\alpha=1$ three-body resonance} \label{app:3br}
\begin{figure*} 
\centering
\includegraphics[width=0.9\linewidth]{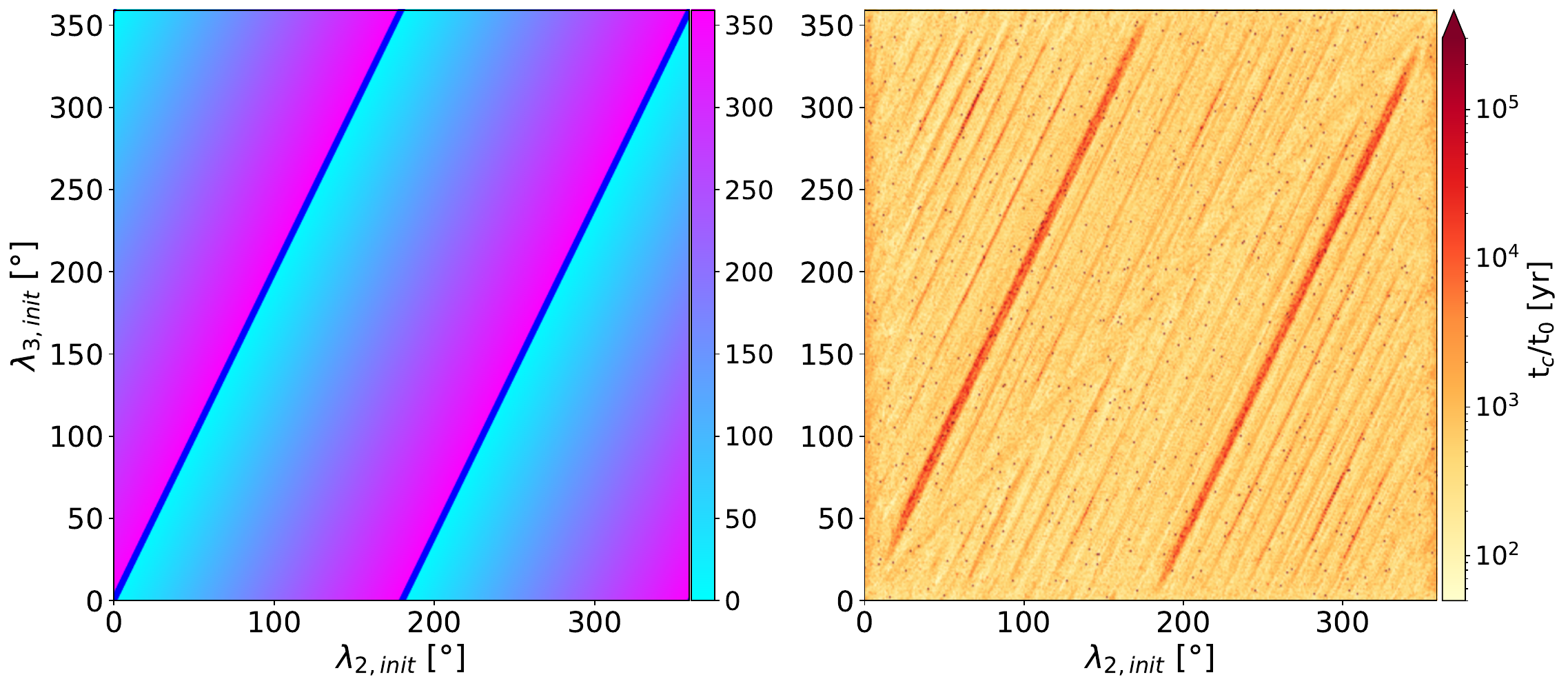}
\caption{Left: resonant angle $\phi$ with $p=1$ and $q=1$ for all values of the angles $\lambda_2$ and $\lambda_3$. The blue stripes correspond to $\phi=0°$. Right: Same as Fig.\,\ref{fig:stripes54135}, but for the system $P_1/P_2 = 0.930$ and $P_2/P_3 = 0.925$ that starts on top of the 3BR. \label{fig:super_3br} }
\end{figure*}

Due to the fact that the systems in SPK1, SPK2, and SPK3 all start on the main diagonal (away from 3BR$\alpha1$), it is not possible to clearly identify the resonance patterns as a function of the initial longitudes in Fig.\,\ref{fig:stripes382086}. We define a system with the same inner period ratio as $b382$, with the outer period ratio such that the system starts on top of 3BR$\alpha1$ (using Eq.\,\ref{eq:pratio}), instead of having the same period ratio value for each pair. The result is shown in Fig. \ref{fig:super_3br}. The "long-lived" stripes correspond to the resonant angle $\phi = -2 \lambda_2 + \lambda_3 = 0$°. On the stripes, 2016, 612, and 6 systems survived for at least 10$^4$, 10$^5$, and 10$^6$, respectively. These systems are initially outside a long-lived island (the one on top of the $14/13$ and $13/12$ 2BRs) and are extremely tightly packed, as the inverse initial inner period ratio represents $\sim 91\%$ of the Hill stability limit value.

\section{More about the geometry of three-body resonances} \label{app:geo}

In this section we turn off gravitational interactions between planets such that the resonances become infinitely thin. We also consider only coplanar and circular orbits.

Let us introduce a time-dependent parameter,

\begin{equation}
\label{eq:gamma}
       \gamma(t) = \Sigma^{N_p-1}_{i=1} \cos\big[\lambda_{i+1}(t) - \lambda_{i}(t)\big],
\end{equation}

\noindent where $N_\mathrm{p}$ is the number of planets in the system. This parameter $\gamma$ represents the sum of the difference in angle between adjacent pairs of planets and, therefore, characterizes the successive conjunctions as a function of time. When $\gamma(t) = N_\mathrm{p} - 1$, all planets are in perfect conjunction. When $\gamma(t) = 1 - N_\mathrm{p}$, then all consecutive pairs of planets are in opposite phase. 

In the case of a three-planet system, Eq.~\ref{eq:gamma} reads

\begin{equation}
\label{eq:gamma2}
       \gamma(t) =  \cos\big[\lambda_{2}(t) - \lambda_{1}(t))\big] + \cos\big[(\lambda_{3}(t) - \lambda_{2}(t)\big].
\end{equation}

\noindent When $\gamma = 0$, then the sum of the difference in longitude between each consecutive planet is 180°, so when a conjunction of two planets occurs, the other planet is in opposite phase. 
The resonance 3BR$\alpha$1 is the only resonance where $\gamma$ can remain constantly equal to zero, in this case during a two-planet conjunction, the other planet is in opposite phase.

Considering $\lambda_i = n_i t + \lambda_{i,init}$ and noting $\omega_{ji} = n_j - n_i$, Eq.~\ref{eq:gamma2} becomes

\begin{equation}
\label{eq:gamma3}
       \gamma(t) =  \cos\big[\omega_{21}t + \lambda_{2,init}\big] + \cos\big[\omega_{32}t + (\lambda_{3,init} - \lambda_{2,init})\big],
\end{equation}


\noindent where $\lambda_{i,init}$ is the initial angle of the $ith$ planet at the start of integration (note that $\lambda_{1,init} = 0$). This represents the sum of two periodic waves with different frequencies and phases. The sum of two cosine can be written as the product of two cosine such that

\begin{equation}
\label{eq:gamma4}
       \gamma(t) =  2 A \times B
,\end{equation}

\noindent where

\begin{equation}
\label{eq:amplitude}
       A \equiv   \cos \bigg[ \bigg( \frac{\omega_{21}-\omega_{32}}{2}\bigg) t +     \bigg( \frac{2 \lambda_{2,init}-\lambda_{3,init}}{2}\bigg) \bigg] ,
\end{equation}

\noindent and 

\begin{equation}
\label{eq:module}
       B \equiv  \cos \bigg[ \bigg( \frac{\omega_{21}+\omega_{32}}{2}\bigg) t +      \frac{\lambda_{3,init}}{2} \bigg].
\end{equation}

\noindent The time-dependent term in Eq.\,\ref{eq:amplitude} can be written as

\begin{equation}
\label{eq:timedep}
       \frac{\omega_{21}-\omega_{32}}{2} = -\frac{1}{2} \bigg[ n_1 - 2 n_2 + n_3 \bigg].
\end{equation}

\noindent In the case where the system is captured in a 3BR, the term inside the brackets can be written as

\begin{equation}
\label{eq:timedep2}
    n_1 - 2 n_2 + n_3 = \frac{q-p}{p} \bigg(n_2 - n_3 \bigg).
\end{equation}

\noindent If the system is captured in the 3BR$\alpha1$, the time-dependent term in Eq.\,\ref{eq:amplitude} is equal to zero. The parameter $\gamma$ is therefore a periodic function with a constant amplitude $A$ that depends only on the phase differential of the outer two planets $2 \lambda_{2,init} - \lambda_{3,init}$. The amplitude is $A=0$ when the phase differential is $\pm \pi$ (when $\phi = \pi$). The third planet is exactly in opposite phase during all successive two-planet conjunctions. On the other hand, if $\phi = 2\pi$, then the amplitude $A=2$ and a perfect three-planet conjunction occurs. The successive configurations are repeated systematically $p$ times during the resonant period $P_{3br}$. The 3BR$\alpha1$ is the only three-body resonance of the network with constant amplitude. For this specific resonance, the trajectory of cos($\lambda_3$ - $\lambda_2$) versus cos($\lambda_2$ - $\lambda_1$) is an ellipse in phase space, whose eccentricity depends on the longitude shift. If the resonant angle is $\phi=\pi/2$, then the trajectory is a circle of radius 1, whereas if $\phi=2\pi$ or $\pi$, the trajectory is a positive or negative diagonal line, respectively.

All 3BRs with the same $\alpha$ value, i.e., all resonances such that $q/p = q_l/p_l$ where $p_l$ and $q_l$ are the lowest terms, will show the same periodic curve of $\gamma$; however, the period $P_{3br}$ increases by a factor $N$, where $N \in 1,2,...$ is the greatest common factor for $p$ and $q$. The number of times the pattern repeats during $P_{3br}$ therefore increases by $N$. This means that the lowest terms define the geometry of the resonance since all resonances with the same $\alpha$ value have the same trajectory in phase space (only the period changes). If $q_l - p_l$ is an even number, then the behavior of $\gamma$ during the second half of the period $P_{3br}$ is the symmetry of the first half along the time axis:

\begin{equation}
\gamma(t > P_{3br}/2) = -\gamma(t < P_{3br}/2), \forall t \in [0, P_{3br}]. 
\end{equation}

\noindent This illustrates, similarly to second-order 2BRs, that the system goes back to the initial configuration in opposite phase (all initial angles are shifted by $180^\circ$ in inertial space) after a period $P_{3br}/2N$.

\begin{figure*} 
\centering
\includegraphics[width=0.9\linewidth]{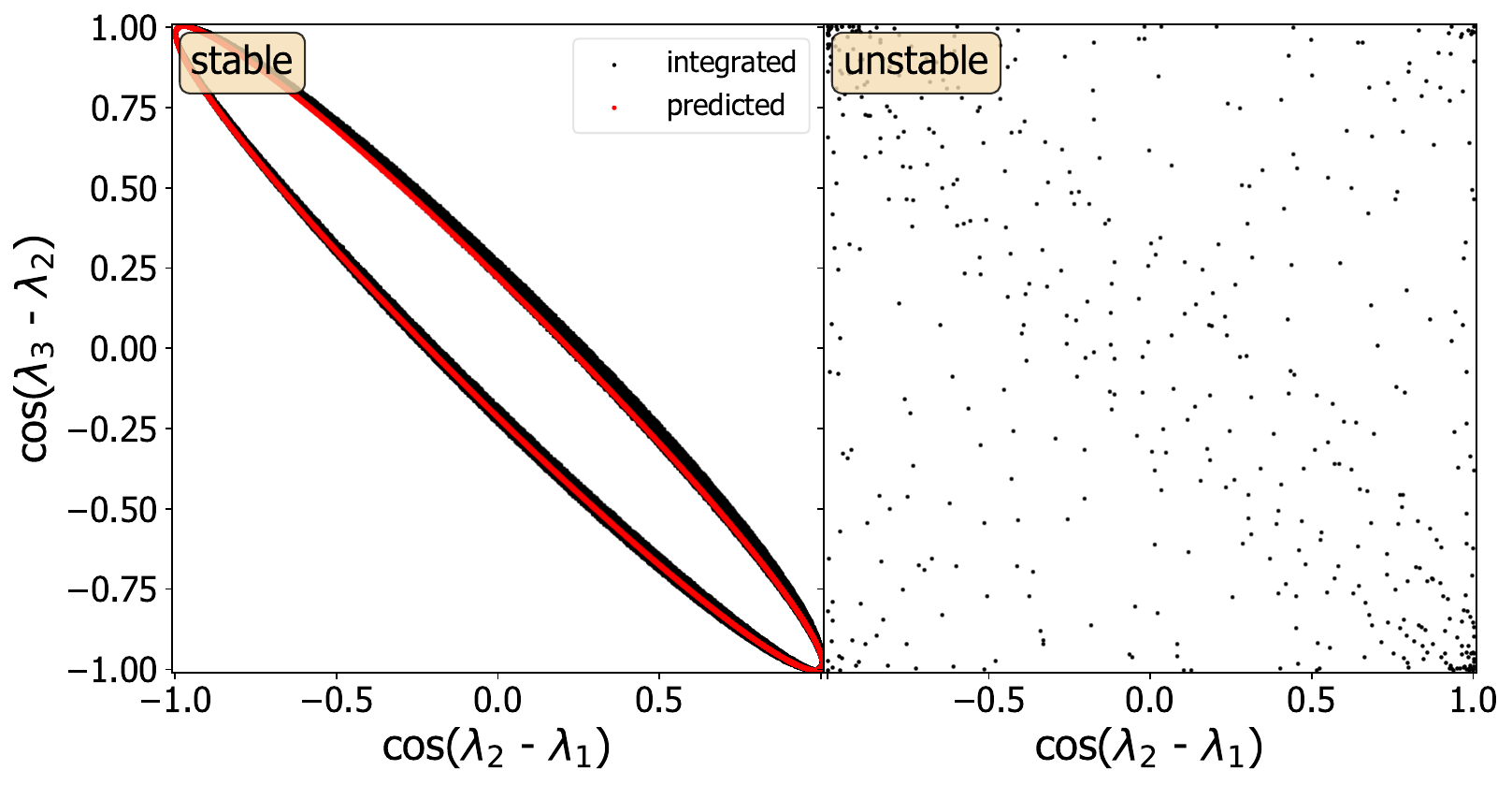}
\caption{Phase space diagram of $\gamma$ for two systems with the same initial $\beta = 3.82086$, but each with a different initial set of longitudes. The left panel shows $\gamma$ for the stable system $b382$ (Specific Set \#1) and the right panel for a chosen set of initial angles ($\lambda_2 = 155$° and $\lambda_3 = 50$°) that quickly leads to chaotic regime. In the right panel, the trajectory is shown for the first 500 years, which roughly corresponds to the unstable system's lifetime, while it is shown for $10^{8}$ years of integration for the stable system. In the left panel, the red line is the predicted trajectory in phase space using Eq.\,\ref{eq:gamma3} (when the planet-planet gravitational interaction is turned off) with the same initial parameters as the stable integrated system. \label{fig:gamma382} }
\end{figure*}

Let us now turn on gravitational interaction between planets. Figure \ref{fig:gamma382} displays the trajectory of $\gamma$ in phase space for the numerically integrated system $b382$. We compare two sets of initial longitudes: one that keeps the system stable (from the Specific Set S1) and another that leads to quick instability. The stable integrated system follows the pattern predicted for a system in the 3BR$\alpha1$ (see the red curve) as it is effectively trapped by this resonance (Fig.\,\ref{fig:res_width}). We can see that the value of $\gamma$ is always well below one, which means that the system always remains very far from a three-planet close approach. On the other hand, the value of $\gamma$ in the unstable system does not follow the resonant pattern and exhibits multiple three-planet close approaches. 

\begin{figure*} 
\centering
\includegraphics[width=0.9\linewidth]{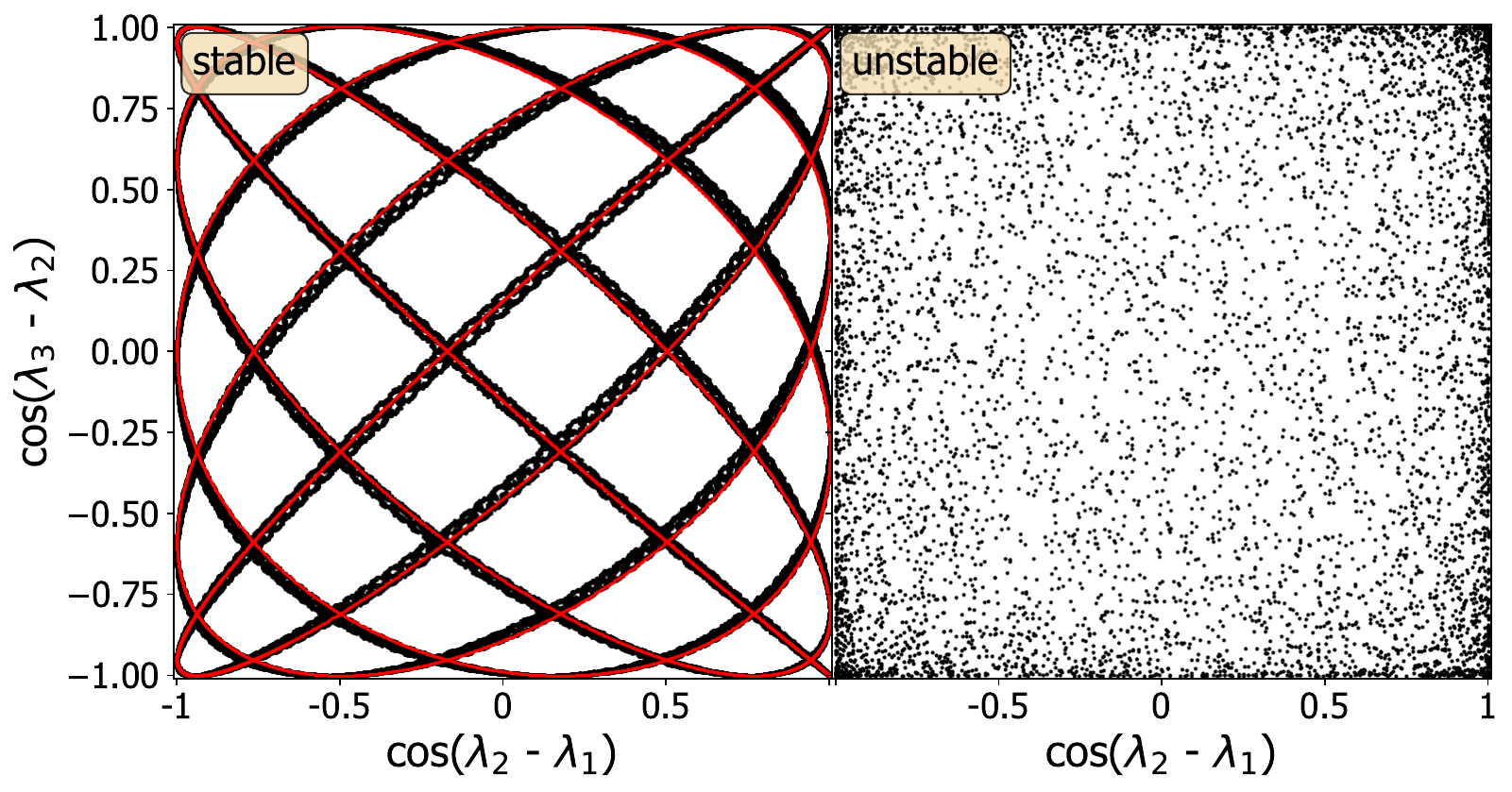}
\caption{Phase space diagram showing the trajectory of $\gamma$ of two systems with the same initial $\beta = 5.4135$ (with $\alpha=10/9$), but each with a different initial set of longitudes. The left panel shows $\gamma$ for the stable configuration $b541$ ($\lambda_2 = 228.2280$° and $\lambda_3 = 145.7270$°) and the right panel an unstable configuration ($\lambda_2 = 230$° and $\lambda_3 = 65$°). For the unstable systems, the initial angle values are chosen so they are away from the black stripes in Fig.\,\ref{fig:stripes54135}. This system became unstable after about 24,000 years but the trajectory is shown during the first 500 years for visibility. The trajectory is shown for $10^{8}$ years of integration for the stable system. The red line is the predicted trajectory using Eq.\,\ref{eq:gamma3} (when the planet-planet gravitational interaction is turned off) with the same initial parameters as the integrated system. \label{fig:gamma541} }
\end{figure*}

The result shows that the trajectory of $\gamma$ in the stable system is equivalent to that of a system with no planet-planet interaction; the net gravitational energy is perfectly balanced during the resonant period. We also examine system $b541$ with the initial longitudes (Sect.\,\ref{sec:spikes45}) and compare it to a different initial set of longitudes that leads to instability (Fig.\,\ref{fig:gamma541}). Similarly, the stable system shows a pattern of $\gamma$ that remains purely shaped by the 3BR$\alpha\frac{10}{9}$.

\section{Two-body resonance angles} \label{app:2br}

\begin{figure} 
\centering
\includegraphics[width=0.9\linewidth]{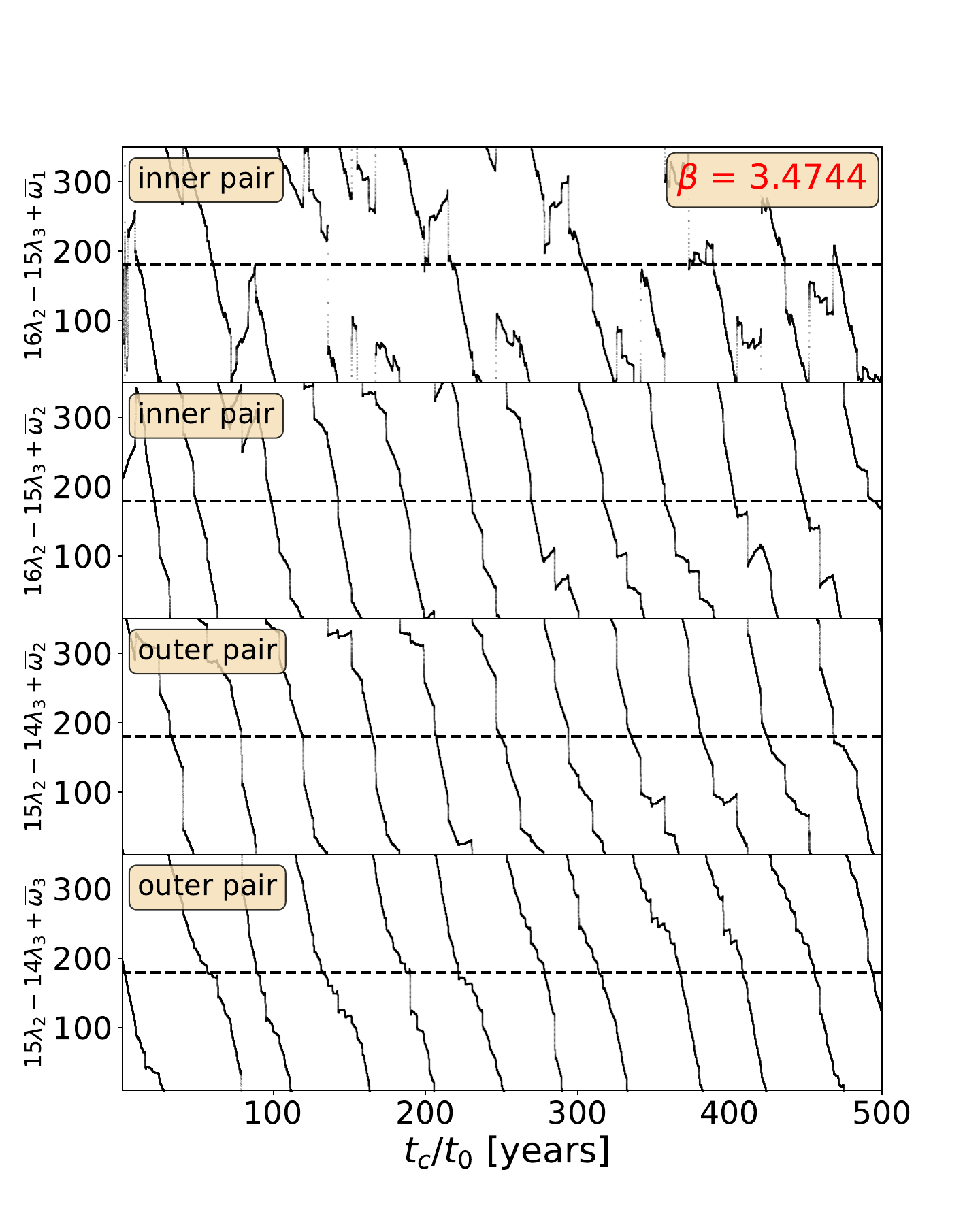}
\caption{Evolution of the first-order two-body resonant angles, 14:15 and 15:16, over the first 1000 orbits of system $\beta=3.4744$ in Set S2. Every angle circulates.   \label{fig:2b_347} }
\end{figure}

\begin{figure} 
\centering
\includegraphics[width=0.9\linewidth]{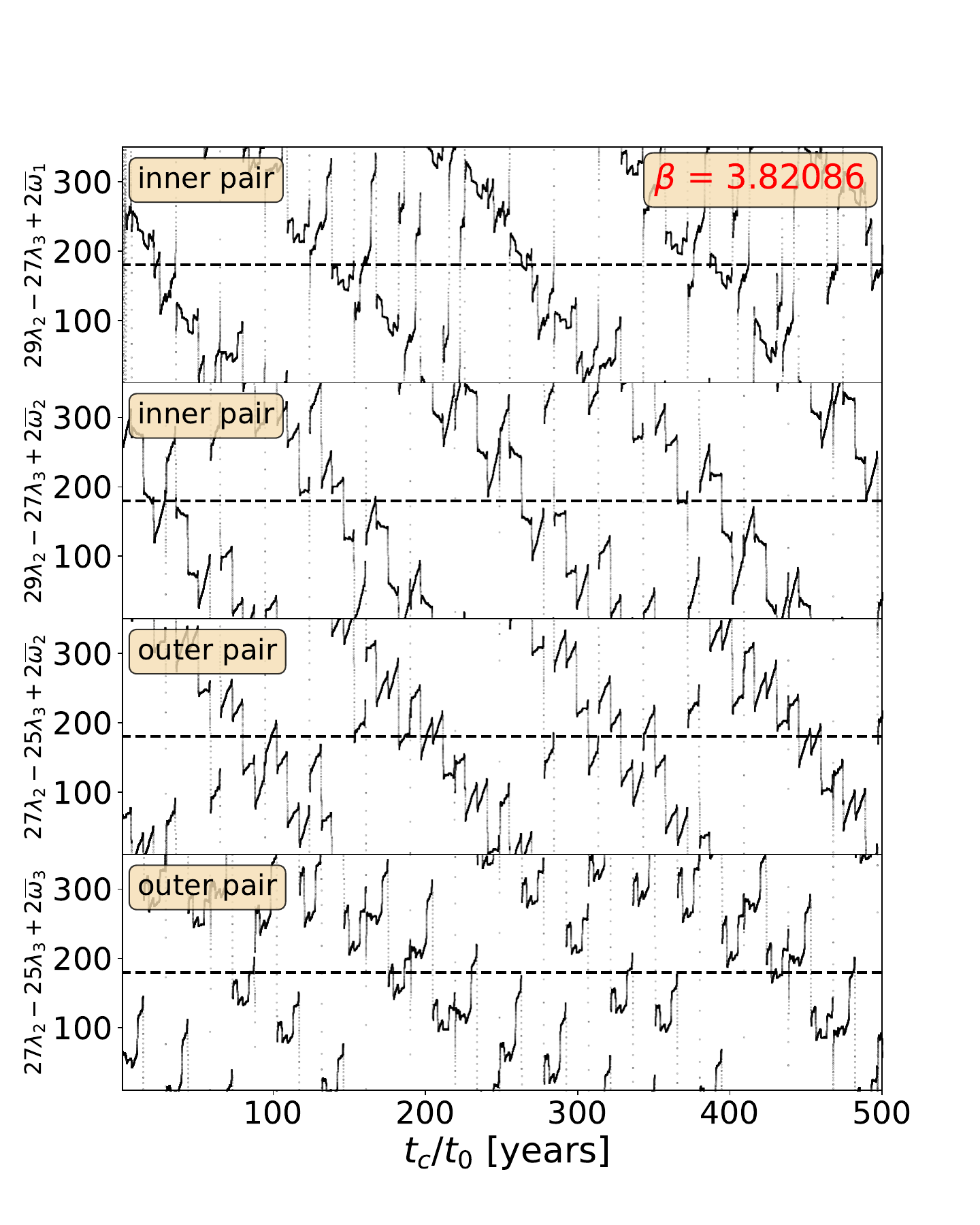}
\caption{Evolution of the second-order two-body resonant angles, 25:27 and 27:29, over the first 1000 orbits of system $\beta=3.82086$ in SPK1. Every angle circulates. \label{fig:2b_382} }
\end{figure}

\begin{figure} 
\centering
\includegraphics[width=0.9\linewidth]{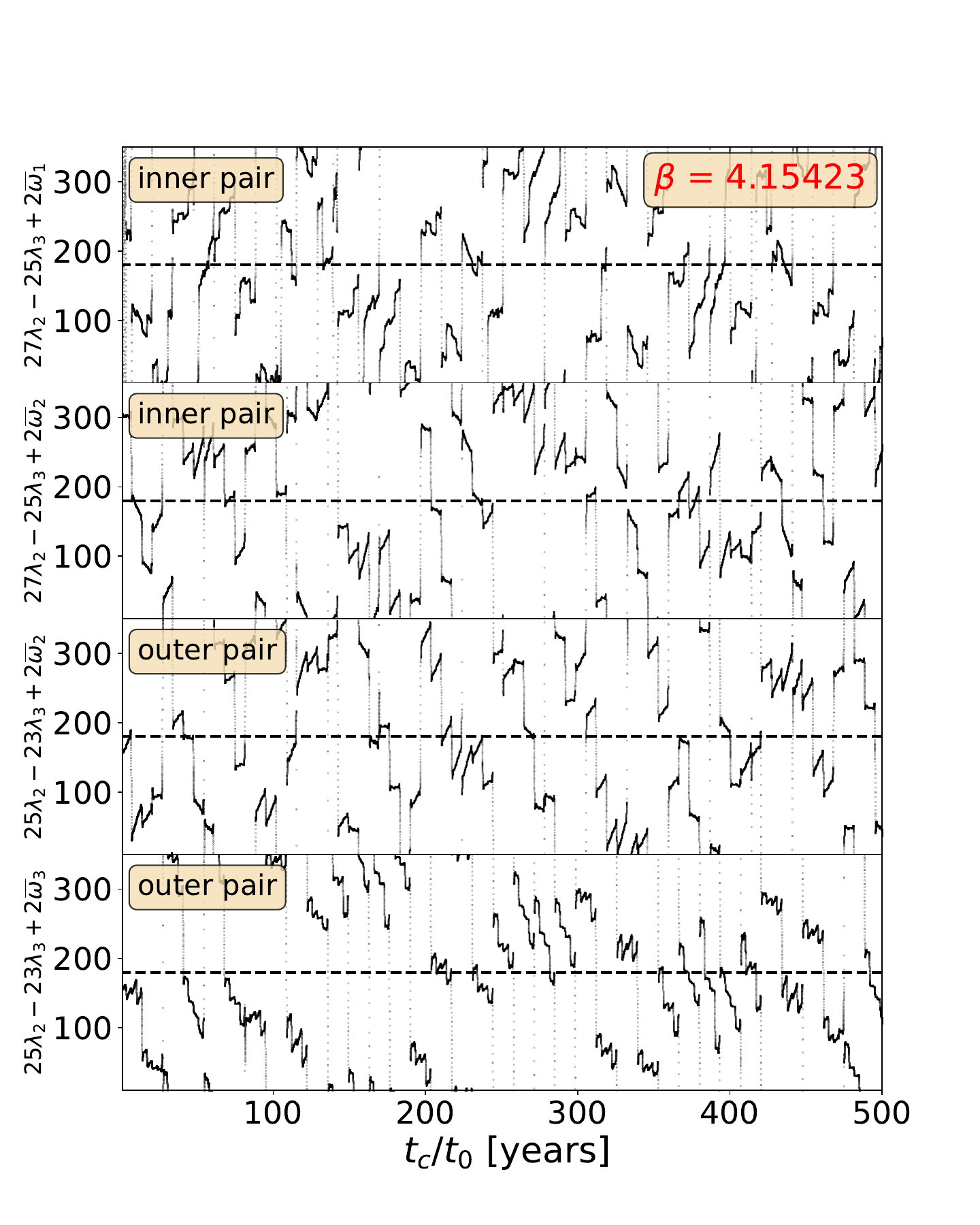}
\caption{Evolution of the second-order two-body resonant angles, 23:25 and 25:27, over the first 1000 orbits of system $\beta=4.15423$ in SPK2. The angles seem to circulate.  \label{fig:2b_415} }
\end{figure}

\begin{figure} 
\centering
\includegraphics[width=0.9\linewidth]{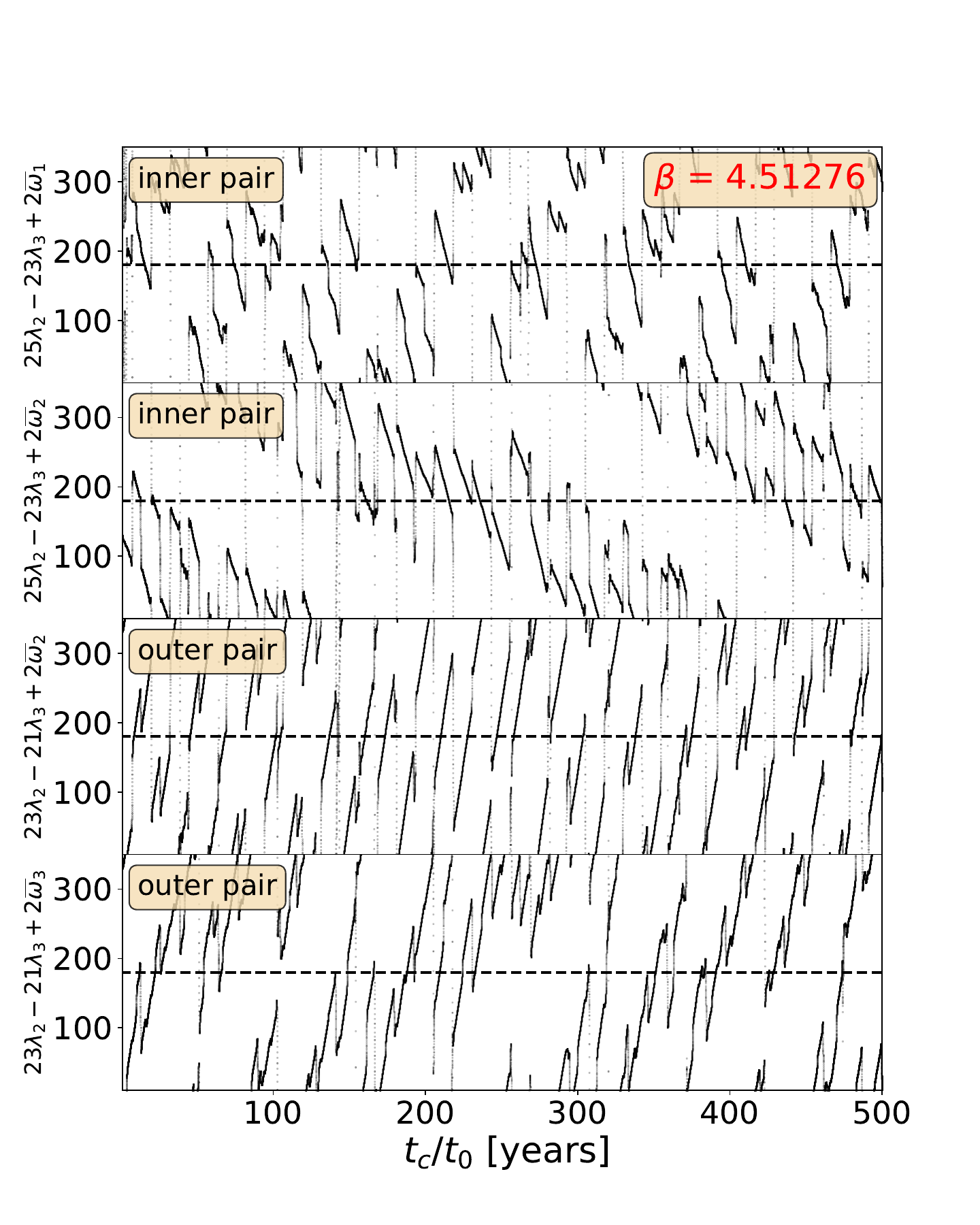}
\caption{Evolution of the two-body resonant angles, 21:23 and 23:25, over the first 1000 orbits of system $\beta=4.51276$ in SPK3. Every angle circulates. \label{fig:2b_451} }
\end{figure}

Figures \ref{fig:2b_347}, \ref{fig:2b_382}, \ref{fig:2b_415}, and \ref{fig:2b_451} show the early evolution of the two-body resonance angles in the direct vicinity of long-lived systems $b347$, $b382$, $b415$, and $b451$, respectively. Almost all angles circulate, and none of them clearly librates, even within a limited range in time. It is not possible to conclude that these systems are also locked in two-body mean motion resonance in addition to three-body resonance. 

\end{appendix}

\end{document}